\documentclass[12pt]{article}
\usepackage{amssymb}
\usepackage{amsmath}
\usepackage{latexsym}
\usepackage{array}
\usepackage{multirow}
\usepackage{epsfig}
\usepackage{rotating}
\usepackage{graphicx}
\usepackage{subfigure}
\usepackage{wasysym}
\usepackage{threeparttable}
\usepackage{lscape}
\usepackage{natbib}
\usepackage{color}
\usepackage{hyperref}
\usepackage{epstopdf}
\usepackage{booktabs}
\usepackage{algorithm}
\usepackage{algorithmicx}

\newcommand{\iid}{\stackrel{\mathrm{iid}}{\sim}}
\newcommand{\Lagr}{\mathcal{L}}

\newcommand{\Appendix}
{
\def\thesection{Appendix~\Alph{section}}
\def\thesubsection{\Alph{section}.\arabic{subsection}}
\def\thesubsection{A.\arabic{subsection}}
}
\def\boxit#1{\vbox{\hrule\hbox{\vrule\kern6pt
          \vbox{\kern6pt#1\kern6pt}\kern6pt\vrule}\hrule}}
\begin{document}
\thispagestyle{empty} \baselineskip=28pt

\begin{center}
{\LARGE{\bf Bayesian Lattice Filters for Time-Varying Autoregression and Time-Frequency Analysis}}
\end{center}

\baselineskip=12pt

\vskip 2mm
\begin{center}
Wen-Hsi Yang\footnote{\baselineskip=12pt CSIRO Computational Informatics,
Ecosciences Precinct, GPO Box 2583, Brisbane QLD 4001, Australia, Wen-Hsi.Yang@csiro.au},
Scott H. Holan\footnote{\baselineskip=12pt
(To whom correspondence should be addressed) Department of Statistics, University of Missouri,
146 Middlebush Hall, Columbia, MO 65211-6100, holans@missouri.edu},
Christopher K. Wikle\footnote{\baselineskip=12pt Department of Statistics, University of Missouri,
146 Middlebush Hall, Columbia, MO 65211-6100, wiklec@missouri.edu}  
\\
\end{center}
%
%
%
%
\vskip 4mm

\begin{center}
{\bf Abstract} 
\end{center}
Modeling nonstationary processes is of paramount importance to many scientific disciplines including environmental science, ecology, and finance, among others.  Consequently, flexible methodology that provides accurate estimation across a wide range of processes is a subject of ongoing interest.  We propose a novel approach to model-based time-frequency estimation using time-varying autoregressive models. In this context, we take a fully Bayesian approach and allow both the autoregressive coefficients and innovation variance to vary over time.  Importantly, our estimation method uses the lattice filter and is cast within the partial autocorrelation domain.  The marginal posterior distributions are of standard form and, as a convenient by-product of our estimation method, our approach avoids undesirable matrix inversions. As such, estimation is extremely computationally efficient and stable.  To illustrate the effectiveness of our approach, we conduct a comprehensive simulation study that compares our method with other competing methods and find that, in most cases, our approach performs superior in terms of average squared error between the estimated and true time-varying spectral density. Lastly, we demonstrate our methodology through three modeling applications; namely, insect communication signals, environmental data (wind components), and macroeconomic data (US gross domestic product (GDP) and consumption).

\baselineskip=12pt

%
%
%

\baselineskip=12pt
\par\vfill\noindent
\par\medskip\noindent
\clearpage\pagebreak\newpage \pagenumbering{arabic}
\baselineskip=24pt
\section{Introduction}\label{intro}
Recent advances in technology have lead to the extensive collection of complex high-frequency nonstationary signals across a wide array of scientific disciplines. In contrast to the time-domain, the time-varying spectrum may provide better insight into important characteristics of the underlying signal \citep[e.g.,][and the references therein]{holan2010modeling, holan2012finance,rosen2012adaptspec,yangwh2013multifun}. For example, \cite{holan2010modeling} demonstrated that features in the time-frequency domain of nonstationary {\it{Enchenopa}} treehopper mating signals may describe crucial phenotypes of sexual selection.

In general, time-frequency analyses can either proceed using a nonparametric or model-based (parametric) approach. The most common nonparametric approach is the short-time Fourier transform (i.e., windowed Fourier transform) which produces a time-frequency representation characterizing local signal properties \citep{grochenig2001foundations, oppenheim2009discrete}.   Another path to time-frequency proceeds using smoothing splines \citep{rosen2009localspectral, rosen2012adaptspec} or by parameterizing the spectral density to estimate the local spectrum via the Whittle likelihood \citep{everitt2013timefrequency}. Similarly, time-frequency can be achieved by applying smooth localized complex exponential (SLEX) functions to the observed signal \citep{ombao2001auto}.  In contrast to window based approaches, the SLEX functions are produced using a projection operator and are, thus, simultaneously orthogonal and localized in both time and frequency.   Alternatively, one can use the theory of frames and over-complete bases to produce a time-frequency representation. For example, continuous wavelet transforms \citep{vidakovic1999statistical, percival2000wavelet, mallat2008wavelet} or Gabor frames \citep{wolfe2004bayesian, feichtinger1998gabor, fitzgerald2000nonlinear} could be used. By introducing redundancy into the basis functions, these representations may provide better simultaneous resolution over both time and frequency.       

Model-based approaches typically proceed through the time-domain in order to produce a time-frequency representation for a given nonstationary signal. In this setting common approaches include fitting piecewise autoregressive (AR) models as well as time-varying autoregressive (TVAR) models. The former approach assumes that the nonstationary signal is piecewise stationary. Consequently, the estimation procedure attempts to identify the order of the AR models along with the location of each piecewise stationary series. For example, \cite{davis2006break} propose the AutoPARM method using minimum description length (MDL) in conjunction with a genetic algorithm (GA) to automatically locate the break points and AR model order within each segmentation. In addition to providing a time-frequency representation, this approach also locates changepoints. \cite{wood2011bayesianmixtures} propose fitting mixtures of AR models within each segment via Markov chain Monte Carlo (MCMC) methods. Their approach selects a common segment length and then divides the signal into these segments prior to implementation of the fitting procedure. Although such approaches may accommodate signals with several piecewise stationary structures, they lack the capability of capturing momentary shocks to the system (i.e., changes to the evolutionary structure that only occur over relatively few time points).   

For many processes, TVAR models may provide superior resolution within the time-frequency domain for both large and small scale features through modeling time-varying parameters. To estimate the TVAR model coefficients, \cite{kitagawa1996smoothness} and \cite{kitagawa2010timeseries} treat the coefficients as a stochastic process and model them using difference equations under the assumption of a maximum fixed order of the TVAR model. Their estimation procedure for the coefficients is then based on state-space models with smoothness priors. In this context, the innovation variances are treated as constant and estimated using a maximum likelihood approach. Subsequently, \cite{west1999evaluation} propose a fully Bayesian TVAR framework that simultaneous models the coefficients and the innovation variances using random walk models. Alternatively, by assuming a constant innovation variance, \cite{prado2002order} model the coefficients and order of the TVAR model using random walk models. Further, to make the TVAR models stable, the constraint that the roots of the characteristic polynomial lie within the unit circle could be imposed. However, such an added condition makes estimation more complicated and computationally expensive.

To avoid these issues, we instead work with the partial autocorrelation coefficients (i.e., in the partial correlation (PARCOR) coefficient domain) and then use the Levinson recursion to connect the PARCOR coefficients and TVAR model coefficients \citep{kitagawa1996smoothness, godsill2004mcsmoothing}. \cite{godsill2004mcsmoothing} model the PARCOR coefficients and innovation standard deviations using a truncated normal first-order autoregression and a log Gaussian first-order autoregression, respectively, with a given constant order. To estimate these values, a sequential Monte Carlo algorithm is used. Alternatively, as previously alluded to, by assuming a constant innovation variance, \cite{kitagawa1996smoothness} implement the smoothness prior within a lattice filter to estimate the PARCOR coefficients. After the PARCOR coefficients have been estimated, a constant innovation variance is estimated using a maximum likelihood approach. However, the former approach is computationally expensive and may suffer from the degeneracy problem (i.e., the collapse of approximations of the marginal distributions) when TVAR model order is large. In addition, certain hyperparameter values (i.e., the TVAR coefficients associated with the two latent models) may be sensitive to starting values and may require prior knowledge or expert supplied subjective information to achieve convergence \citep{godsill2004mcsmoothing}. Although the latter takes advantage of the lattice form to estimate the PARCOR coefficients, estimation of the innovation variance is achieved outside of the lattice structure; that is, the estimation procedure is a two-stage method. Importantly, this approach is designed for a constant innovation variance and cannot deal with time-dependent innovation variances. As such, we propose a novel approach that addresses these issues within a fully Bayesian context. 

\cite{kitagawa1988numeric} and \cite{kitagawa1996smoothness} constitute the first attempts at utilizing the lattice structure to estimate the coefficients of a TVAR model when the innovations follow a Gaussian distribution with mean zero and constant variance. At each stage of the lattice filter, they assume that the residual at each time between the forward and backward prediction errors follows a Cauchy distribution, and that the PARCOR coefficient is modeled as a Gaussian random walk. This produces a non-Gaussian state space model at each stage and thus, a numeric algorithm is conducted for estimation. Moreover, the assumptions of their approach ignore an implicit connection between the innovation term of the TVAR model and the residual term between the forward and backward prediction errors. That is, the distribution of residual term should be a Gaussian distribution rather than a Cauchy distribution. Consequently, their approach leads to a TVAR model with innovation terms following a Cauchy distribution; hence the innovation variances do not exist. In contrast, our approach assumes the residual term follows a Gaussian distribution. 

We propose a fully Bayesian approach to efficiently estimate the TVAR coefficients and innovation variances within the lattice structure. One novel aspect of our approach is that we model both the PARCOR coefficients and the TVAR innovation variances within the lattice structure and then estimate them simultaneously. This is different from the frequentist two-stage method of \cite{kitagawa1988numeric} and \cite{kitagawa1996smoothness}. Another novel aspect is that we take advantage of dynamic linear model (DLM) theory \citep{west1997bayesian, prado2010time} to regularize the PARCOR coefficients instead of using truncated distributions. Thus, our method provides marginal posterior distributions with standard forms for both the PARCOR coefficients and innovation variances. Since our approach takes advantage of the lattice structure, the computational efficiency of our approach is not affected by the order of the TVAR model; that is, our approach avoids having to calculate higher dimensional inverse matrices. To select the TVAR model, we provide both a visual and a numerical method. Importantly, the simulation study we provide demonstrates that our approach leads to superior performance in terms of estimating the time-frequency representation of various nonstationary signals, as measured by average squared error. Thus, our approach provides a stable and computationally efficient way to fit TVAR models for time-frequency analysis. 

The remainder of this paper is organized as follows. Section~\ref{sec:methodology} briefly introduces the lattice structure and describes our methodology along with prior specification. Section~\ref{sec:simulation} presents a comprehensive simulation study that illustrates the effectiveness of our approach across an expansive array of nonstationary processes. Subsequently, in Section~\ref{sec:application}, our methodology is demonstrated through three modeling applications; namely, insect communication signals, environmental data (i.e., wind components), and macroeconomic data (i.e., US gross domestic product (GDP) and consumption).  Lastly, Section~\ref{sec:discussion} concludes with discussion. For convenience of exposition, details surrounding the estimation algorithms and additional figures are left to an Appendix.

\section{Methodology}\label{sec:methodology}
\subsection{Time-Varying Coefficient Autoregressive Models}
The TVAR model of order $P$ for a nonstationary univariate time series $x_{t}$, $t=1,\ldots,T$, can be expressed as
\begin{equation}\label{forward_tvar}
x_{t}=\sum_{m=1}^{P}a_{t,m}^{(P)}x_{t-m}+\epsilon_{t},
\end{equation}
where $a_{t,m}^{(P)}$ and $\epsilon_{t}$ are the TVAR coefficients associated with time lag $m$ at time $t$ and the innovation at time $t$, respectively. Typically, the innovations are assumed to be uncorrelated mean-zero Gaussian random variables (i.e., $\epsilon_{t}\sim N(0,\sigma_{t}^{2})$, with time-varying variance $\sigma_{t}^{2}$). Therefore, the TVAR model corresponds to a nonstationary AR model with the AR coefficients and variances evolving through time. In such settings, the model is locally stationary but nonstationary globally. As will be illustrated, the assumption of local stationarity is not required for our approach; that is, the forward and backward partial autocorrelations (defined in Section~\ref{subsec:lattice}) need not be equal. Because this model generally allows both slow and rapid changes in the parameters, it can flexibly model the stochastic pattern changes often exhibited by complex nonstationary signals.   
  
\subsection{Lattice Structures}\label{subsec:lattice}
The Levinson-Durbin algorithm yields a unique correspondence between the PARCOR coefficients and the AR coefficients  \citep{shumway2006time,kitagawa2010timeseries}.  Therefore, a disciplined approach to fitting AR models can be achieved through estimation of the PARCOR coefficients.  The lattice structure described below provides a direct way of associating the PARCOR coefficients with the observed time series (see \citet[][Page 225]{hayes1996sdspam} and the Supplementary Appendix for additional discussion).  As such, the lattice structure provides an effective path to AR model estimation.  In fact, the Levinson-Durbin algorithm for a stationary time series can be derived using the lattice structure \citep[see][Appendix~B]{kitagawa2010timeseries}.

Let $f_{t}^{(P)}$ and $b_{t}^{(P)}$ denote the prediction error at time $t$ for a forward and backward AR($P$) model, respectively, where
\begin{eqnarray*}
f_{t}^{(P)}&=x_{t}-\sum_{m=1}^{P}a_{m}^{(P)}x_{t-m}\,\,\,\,\,\mbox{and}\,\,\,\,\, b_{t}^{(P)}&=x_{t}- \sum_{m=1}^{P}d_{m}^{(P)}x_{t+m}.
\end{eqnarray*}
Then, the $m$-th stage of the lattice filter can be characterized by the pair of input-output relations between the forward and backward predictions,
\begin{align}\label{stat_lattice1}
f_{t}^{(m-1)}&=\alpha_{m}^{(m)}b_{t-m}^{(m-1)} + f_{t}^{(m)} ,\\
b_{t}^{(m-1)}&=\beta_{m}^{(m)}f_{t+m}^{(m-1)} + b_{t}^{(m)},~~~~m=1,2,\ldots,P,
\label{stat_lattice2}
\end{align}
with the initial condition, $f_{t}^{(0)}=b_{t}^{(0)}=x_{t}$, and where $\alpha_{m}^{(m)}$ and $\beta_{m}^{(m)}$ are the lag $m$ forward and backward PARCOR coefficients, respectively. Equation \eqref{stat_lattice1} shows that the forward PARCOR coefficient of lag $m$ is a regression coefficient of the forward prediction error $f_{t}^{(m-1)}$ regressed on the backward prediction error $b_{t-m}^{(m-1)}$ and the residual term $f_{t}^{(m)}$ is the forward prediction error of the forward AR($m$) model. Similarly, \eqref{stat_lattice2} shows that the backward PARCOR coefficient of lag $m$ is a regression coefficient of the backward prediction error $b_{t}^{(m-1)}$ regressed on the forward prediction error $f_{t+m}^{(m-1)}$ and the residual term $b_{t}^{(m)}$ is the backward prediction error of the backward AR($m$) model. Using \eqref{stat_lattice1} and \eqref{stat_lattice2} recursively, we can derive the PARCOR coefficients for a given lag. In the stationary case, the forward and backward PARCOR coefficients are equivalent; i.e., $\alpha_{m}^{(m)} = \beta_{m}^{(m)}$. 

\paragraph{Example:} To illustrate $f_{t}^{(m)}$ and $b_{t}^{(m)}$ on the right hand side of \eqref{stat_lattice1} and \eqref{stat_lattice2} are the prediction errors of the forward and backward AR models, respectively, we consider an example when $P=2$. In this case, we first derive that the difference between $f_{t}^{(1)}$ and $\alpha_{2}^{(2)}b_{t-2}^{(1)}$ of \eqref{stat_lattice1} is equal to the forward prediction errors of an AR(2) model as follows  
\begin{align*}
f_{t}^{(1)} - \alpha_{2}^{(2)}b_{t-2}^{(1)} & =f_{t}^{(0)}-\alpha_{1}^{(1)}b_{t-1}^{(0)}-\alpha_{2}^{(2)}(b_{t-2}^{(0)}-\beta_{1}^{(1)}f_{t-1}^{(0)})
 = x_{t}-\alpha_{1}^{(1)}x_{t-1}-\alpha_{2}^{(2)}(x_{t-2}-\beta_{1}^{(1)}x_{t-1})\\
& = x_{t}- (\alpha_{1}^{(1)}-\alpha_{2}^{(2)}\beta_{1}^{(1)})x_{t-1}-\alpha_{2}^{(2)}x_{t-2}  = f_{t}^{(2)}.
\end{align*}
The above derivation shows that the difference between $f_{t}^{(1)}$ and $\alpha_{2}^{(2)}b_{t-2}^{(1)}$ is equal to the prediction errors of the forward AR(2). It can also be shown that \eqref{stat_lattice2} is true. Moreover, when the signal is stationary, the forward and backward PARCOR coefficients are equal (i.e., $\alpha_{m}^{(m)}=\beta_{m}^{(m)}$). In such cases, we can change the second term of the third row to $(\alpha_{1}^{(1)}-\alpha_{2}^{(2)}\alpha_{1}^{(1)})x_{t-1}$.   
  
The PARCOR coefficients $\alpha_{P}^{(P)}$ are equal to the last component of the coefficients of the forward AR($P$) model; i.e., $\alpha_{P}^{(P)} = a_{P}^{(P)}$. Using the Levinson-Durbin algorithm, the remainder of the AR coefficients and the innovation variance can be obtained as follows
\begin{equation}
a_{m}^{(P)} =a_{m}^{(P-1)}-a_{P}^{(P)}a_{P-m}^{(P-1)},~~~~m=1,\ldots,P-1.
\label{eq:levinson}
\end{equation}
This equation implies that once the PARCOR coefficient $\alpha_{P}^{(P)}$ is estimated, then all of the other coefficients are immediately determined as well. 

\subsection{The Lattice Structure of the TVAR model}\label{sec:latticeTVAR}    
Given the assumption of second-order stationarity, the forward and backward PARCOR coefficients are constant over time (i.e., shift-invariant). However, since most real-world signals are nonstationary, the shift-invariant PARCOR coefficients are typically inappropriate. In such cases we can modify \eqref{stat_lattice1} and \eqref{stat_lattice2} as follows
\begin{align}\label{tvar_lattice1}
f_{t}^{(m-1)}&=\alpha_{t,m}^{(m)}b_{t-m}^{(m-1)} + f_{t}^{(m)} ,\\
b_{t}^{(m-1)}&=\beta_{t,m}^{(m)}f_{t+m}^{(m-1)} + b_{t}^{(m)},~~~~m=1,2,\ldots,P,
\label{tvar_lattice2}
\end{align}
with both the forward and backward PARCOR coefficients $\alpha_{t,m}^{(m)}$ and $\beta_{t,m}^{(m)}$ now time dependent. Note that for notational simplicity, $f_{t}^{(m-1)}$ and $b_{t}^{(m-1)}$ here denote the prediction error at time $t$ of the forward and backward TVAR($m-1$). For locally stationary signals, we may impose the constraint that $\alpha_{t,m}^{(m)} = \beta_{t,m}^{(m)}$ at each time $t$. However, for general nonstationary cases, $\alpha_{t,m}^{(m)}$ and $\beta_{t,m}^{(m)}$ may not be identical at each time $t$. Therefore, our approach will proceed without this constraint. Also, the residual terms, $f_{t}^{(m)}$ and $b_{t}^{(m)}$, are assumed to follow zero-mean Gaussian distributions, $N(0,\sigma_{f,m,t}^{2})$ and $N(0,\sigma_{b,m,t}^{2})$, respectively. Importantly, when the true process is TVAR($P$), the variance $\sigma_{f,P,t}^{2}$ is equal to the innovation variance $\sigma_{t}^{2}$. To verify this statement, we use the fact that PARCOR coefficient of AR($P$) is equal to zero when the lag is larger than $P$. This property can be applied to TVAR models since TVAR models correspond to AR models at each time $t$. Using such property, $\alpha_{t,m}^{(m)}$ of \eqref{tvar_lattice1} is equal to zero for $m>P$. Consequently, $f_{t}^{(m)}=f_{t}^{(m+1)}$ for $m\geq P$. The forward prediction error $f_{t}^{(P)}$ is identical to $\epsilon_{t}$; i.e., $f_{t}^{(P)}$ and $\epsilon_{t}$ are identically distributed.  Therefore, as mentioned in Section~\ref{intro}, the Gaussian distribution provides a more reasonable assumption for the target model \eqref{forward_tvar} than the Cauchy distributions used by \cite{kitagawa1988numeric} and \cite{kitagawa1996smoothness}. 

For each stage $m$ of the lattice structure, we construct the following equations to obtain the coefficients, $a_{t,k}^{(m)}$ and $d_{t,k}^{(m)}$, of the forward and backward TVAR models \citep{hayes1996sdspam,kitagawa1996smoothness,haykin2002adaptive} 
\begin{align}\label{tvar_levison1}
	a_{t,k}^{(m)} &=a_{t,k}^{(m-1)}-a_{t,m}^{(m)}d_{t,m-k}^{(m-1)}\\
			 d_{t,k}^{(m)} &=d_{t,k}^{(m-1)}-d_{t,k}^{(m)}a_{t,m-k}^{(m-1)},~~~~k=1,2,\ldots,m-1,
\label{tvar_levison2}
\end{align}
with $a_{t,m}^{(m)}=\alpha_{t,m}^{(m)}$ and $d_{t,m}^{(m)}=\beta_{t,m}^{(m)}$. Equations \eqref{tvar_levison1} and \eqref{tvar_levison2} describe the relationship between the coefficients of the forward and backward TVAR models. In particular, these relations illustrate that the forward coefficients at the current stage are a linear combination of the forward and backward coefficients of the previous stage, with the weights equal to the PARCOR coefficients. Importantly, such a combination also includes the stationary and locally stationary cases. For the stationary case, since $\alpha_{t,m}^{(m)}=\beta_{t,m}^{(m)}$ are constant over time, the general equations \eqref{tvar_levison1} and \eqref{tvar_levison2} can be reduced to \eqref{eq:levinson}. For locally stationary cases, since $\alpha_{t,m}^{(m)}=\beta_{t,m}^{(m)}$ at time $t$, \eqref{tvar_levison1} and \eqref{tvar_levison2} are identical.      

\subsection{Model Specification and Bayesian Inference}    
Since both the forward and backward PARCOR coefficients of \eqref{tvar_lattice1} and \eqref{tvar_lattice2} as well as the corresponding innovation variances require time-varying structures, we consider random walk models for their evolutions. In such cases, the following two hierarchical components are added to \eqref{tvar_lattice1} and \eqref{tvar_lattice2}.   The evolution of the forward and backward PARCOR coefficients are modeled, respectively, as follows: 
\begin{align}\label{eq:walks_parcor1}
\alpha_{t,m}^{(m)} &= \alpha_{t-1,m}^{(m)} + \epsilon_{\alpha,m,t}, ~~~~ \epsilon_{\alpha,m,t}\sim N(0,w_{\alpha,m,t}),\\
\beta_{t,m}^{(m)} &= \beta_{t-1,m}^{(m)} + \epsilon_{\beta,m,t}, ~~~~ \epsilon_{\beta,m,t}\sim N(0,w_{\beta,m,t}),
\label{eq:walks_parcor2}
\end{align}
where $w_{\alpha,m,t}$ and $w_{\beta,m,t}$ are time dependent system variances. These system variances are then defined in terms of hyperparameters $\gamma_{f,m}$ and $\gamma_{b,m}$, so called {\it{discount factors}} with range $(0,1)$, respectively \citep{west1997bayesian}; see the Appendix for further details. Usually, we treat $\gamma_{f,m}=\gamma_{b,m}=\gamma_{m}$ at each stage $m$. These two equations also imply a sequential update form that the PARCOR coefficient at time $t+1$ is equal to the sum of the PARCOR coefficient at time $t$ plus a correction.  

Similarly, both the evolution innovation variances, $\sigma_{f,m,t}^{2}$ and $\sigma_{b,m,t}^{2}$, are modeled through multiplicative random walks as follows  
\begin{align}\label{eq:walks_innov_var1}
\sigma_{f,m,t}^{2} &= \sigma_{f,m,t-1}^{2}(\delta_{f,m}/\eta_{f,m,t}),~~~~\eta_{f,m,t}\sim Beta(g_{f,m,t},h_{f,m,t}),\\
\sigma_{b,m,t}^{2} &= \sigma_{b,m,t-1}^{2}(\delta_{b,m}/\eta_{b,m,t}),~~~~\eta_{b,m,t}\sim Beta(g_{b,m,t},h_{b,m,t}),
\label{eq:walks_innov_var2}
\end{align}
where $\delta_{f,m}$ and $\delta_{b,m}$ are hyperparameters (i.e., discount factors on the range (0,1)), and the multiplicative innovations, $\eta_{f,t,m}$ and $\eta_{b,t,m}$ follow beta distributions with parameters, $(g_{f,m,t},h_{f,m,t})$ and $(g_{b,m,t},h_{b,m,t})$, respectively \citep{west1999evaluation}. These parameters are defined at each time $t$ by the discount factors, $\delta_{f,m}$ and $\delta_{b,m}$, as detailed in the Appendix. In many cases, we also assume that $\delta_{f,m}=\delta_{b,m}=\delta_{m}$ at each stage $m$. The series of stochastic error terms $\epsilon_{\alpha,m,t}$, $\epsilon_{\beta,m,t}$, $\eta_{f,m,t}$, and $\eta_{b,m,t}$ are mutually independent, and independent of the forward and backward innovations, $f_{t}^{(m)}$ and $b_{t}^{(m)}$ of \eqref{tvar_lattice1} and \eqref{tvar_lattice2}. 

We specify conjugate initial priors for $\alpha_{0,m}^{(m)}$ and $\sigma_{f,m,0}^{2}$ at each stage $m$ as follows 
\begin{align}\label{eq:conjugate1}
p(\alpha_{0,m}^{(m)}|D_{f,m,0},\sigma_{f,m,0}^{2})& \sim N(\mu_{f,m,0},c_{f,m,0}), \\
p(\sigma_{f,m,0}^{-2}|D_{f,m,0}) & \sim G(v_{f,m,0}/2,\kappa_{f,m,0}/2),
\label{eq:conjugate2}        
\end{align}
where $D_{f,m,0}$ denotes the information set at the initial time $t=0$, $G(\cdot,\cdot)$ is the gamma distribution, $\mu_{f,m,0}$ and $c_{f,m,0}$ are the mean and variance for a normal distribution, and $v_{f,m,0}/2$ and $\kappa_{f,m,0}/2$ are the shape and scale parameters for a gamma distribution. Usually, we treat the starting values $\mu_{f,m,0}$, $c_{f,m,0}$, $v_{f,m,0}$, and $\kappa_{f,m,0}$ as common constants over stage $m$. Typically, we choose $\mu_{f,m,0}$ and $c_{f,m,0}$ to be zero and one, respectively.  In addition, to set $v_{f,m,0}$ and $v_{f,m,0}$, we first fix $v_{f,m,0}=1$ and calculate the sample variance of the initial components of the signal. Given these two values, we can obtain $\kappa_{f,m,0}$ through the formula for the expectation of the gamma distribution.  In such prior settings, the DLM sequential filtering and smoothing algorithms provide the necessary components for the marginal posterior distributions \citep{west1997bayesian}.  Specifically, for $t=1,\ldots,T$, with the information set $D_{f,m,T}$ up to time $T$, the marginal posterior distributions $p(\alpha_{t,m}^{(m)}|D_{f,m,T})$ and $p(\sigma_{f,m,t}^{-2}|D_{f,m,T})$ are the $t$-distribution and gamma distribution, respectively.  Analogous to $\alpha_{0,m}^{(m)}$ and $\sigma_{f,m,0}^{2}$, the same conjugate initial priors for $\beta_{0,m}^{(m)}$ and $\sigma_{b,m,0}^{2}$ are specified at each stage $m$. Details of the sequential filtering and smoothing for the PARCOR coefficients and innovation variances for each stage $m$ are discussed in the Appendix.  

\subsection{Model Selection}\label{selection}
Selection of the model order and set of discount factors $\{P,\gamma_{m},\delta_{m}; m=1,\ldots,P \}$ is essential for our approach. First, one can assume $\gamma_{m}=\gamma$ and $\delta_{m}=\delta$, for $m=1,\ldots,P$. Then, the analysis will proceed using a set of various pre-specified combinations of $(P,\gamma,\delta)$. Since $\gamma$ is related to the variability of the PARCOR coefficients, it also affects the variability of the TVAR coefficients.  Hence, one can model the variance of the time-varying coefficients and the innovation variances of the TVAR models using discount factors $\gamma$ and $\delta$, respectively \citep{west1999evaluation}.  Note that, in our context, the discount factors $(\gamma_m, \delta_m)$ are a function of $m$ (the lattice filter stage), whereas the \citet{west1999evaluation} setting does not make use of the lattice filter and, thus, there is only one set of discount factors $(\gamma, \delta)$ that need to be estimated.  However, estimation of $(P, \gamma, \delta)$ using the approach of \citet{west1999evaluation} entails repeatedly having to calculate inverse matrices in the sequential filtering process. Our approach allows $\gamma_{m}$ and $\delta_{m}$ to vary by stage. Thus, we first specify a potential maximum value of $P$ and a set of combinations of $\{\gamma_{m},\delta_{m}\}$ for each stage $m$. Given a value of $P$, we search for the combination of $\{\gamma_{1},\delta_{1}\}$ maximizing the log likelihood of \eqref{tvar_lattice1} at stage one. Using the selected $\gamma_{1}$ and $\delta_{1}$, we can obtain the corresponding series $\{f_{t}^{(2)}\}$ and $\{b_{t}^{(2)}\}$, for $t=1,\ldots,T$, as well as the value, $\Lagr_{1}$, of log maximum likelihood of \eqref{tvar_lattice1}. We then, repeat the above search procedure for stage two using the output $\{f_{t}^{(2)}\}$ and $\{b_{t}^{(2)}\}$ obtained from implementing the selected hyperparameters $\gamma_{1}$ and $\delta_{1}$. In turn, this produces a new series of $\{f_{t}^{(3)}\}$ and $\{b_{t}^{(3)}\}$, for $t=1,\ldots,T$, as well as a value $\Lagr_{2}$. We repeat the procedure until the set of $\{\gamma_{m},\delta_{m},\Lagr_{m}\}$, $m=1,\ldots,P$, has been selected.

Here, we provide both a visual and numerical method to select the order. Similar to the scree plot widely used in multivariate analysis \citep{rencher2002}, we can plot $\Lagr_{m}$ against the order $m$. When the observed series follows an AR or TVAR model, the values of $\Lagr_{m}$ will stop increasing after a specific lag, this lag can be chosen as the order for the estimated model. Henceforth, this plot is referred to as ``BLF-scree." This type of visual order determination can be directly quantified through the relative change of $\Lagr_{m}$. Specifically, we provide a numerical method of order selection based on calculating the percent change in going from $\Lagr_{m-1}$ to $\Lagr_{m}$ with respect to $m$,
\begin{equation}\label{eq:criteria}
|(\Lagr_{m}-\Lagr_{m-1})/\Lagr_{m-1}|*100 < \tau.
\end{equation}

Based on simulation of various TVAR models, we choose $\tau=0.5$ with $m-1$ reflecting the ``best'' value for the order. That is, we have found that $0.5$ provides an effective cut-off for choosing the order.  Although this approach provides a good guide to order selection, other model selection methods could be considered (e.g., shrinkage through Bayesian variable selection, reversible jump MCMC, or by minimizing an information criteria).  Development of alternative model selection approaches in this setting constitutes an area of future research.

We now summarize our approach for fitting TVAR models. Given a set of hyperparameters $\{P,\gamma_{m},\delta_{m}; m=1,\ldots,P \}$, the procedure starts by setting $f_{t}^{(0)}=b_{t}^{(0)}=x_{t}$, for $t=1,\ldots,T$. Next, plugging $\{f_{t}^{(0)}\}$ and $\{b_{t}^{(0)}\}$ into \eqref{tvar_lattice1} and \eqref{tvar_lattice2} and using sequential filtering and smoothing algorithms, we obtain a series of estimated parameters $\{\widehat{\alpha}_{t,1}^{(1)}\}$, $\{\widehat{\beta}_{t,1}^{(1)}\}$, $\{\widehat{\sigma}_{f,1,t}^{2}\}$, and $\{\widehat{\sigma}_{b,1,t}^{2}\}$, as well as the new series of forward and backward prediction errors, $\{f_{t}^{(1)}\}$ and $\{b_{t}^{(1)}\}$, for $t=1\ldots,T$. We then repeat the above procedure until $\{\widehat{\alpha}_{t,P}^{(P)}\}$, $\{\widehat{\beta}_{t,P}^{(P)}\}$, $\{\widehat{\sigma}_{f,P,t}^{2}\}$, and $\{\widehat{\sigma}_{b,P,t}^{2}\}$ have been obtained. Then, recursively plugging the estimates of $\{\alpha_{t,m}^{(m)}\}$ and $\{\beta_{t,m}^{(m)}\}$, from $m=1,\ldots,P$ into \eqref{tvar_levison1} and \eqref{tvar_levison2}, we obtain the estimated time-varying coefficients of \eqref{forward_tvar}. As part of this algorithm, the series of estimated innovation variances are equal to $\{\widehat{\sigma}_{f,P,t}^{2}\}$. Finally, for $t=1,\ldots,T$, the time-frequency representation associated with the TVAR($P$) model can be obtained by the following equation             
\begin{equation}\label{eq:tvspec}
S(t,\omega)= \frac{{\sigma_{t}^{2}}}{\left|1-\sum_{m=1}^{P}a_{t,m}^{(P)}\mbox{exp}(-2\pi i m\omega)\right|^2},~~~~~-1/2 \leq \omega \leq 1/2,
\end{equation}
where $i=\sqrt{-1}$ \citep{kitagawa1996smoothness}. Plugging the estimated values $\widehat{a}_{t,m}^{(P)}$, $m=1,\ldots,P$, and $\widehat{\sigma}_{f,P,t}^{2}$ into \eqref{eq:tvspec} yields the estimated time-varying AR($P$) time-frequency representation $\widehat{S}(t,\omega)$.  See the Appendix for further discussion.

\section{Simulation Studies}\label{sec:simulation}
In this section, we simulate various nonstationary time series in order to compare the performance of our approach with four other approaches used to estimate the time-frequency representation. The first approach is AdaptSPEC proposed by \cite{rosen2012adaptspec}. This approach adaptively segments the signal into finite pieces and then estimates the time-frequency representation using smoothing splines to fit local spectra via the Whittle likelihood approximation. The size of a segment and the number of the spline basis functions are two essential parameters for this approach. To reduce any subjectivity in our comparisons, we choose settings for these two parameters similar to those considered in \cite{rosen2012adaptspec} (with their $t_{min}=40$), as well as the same settings for MCMC iterations and burn-in.  However, rather than using 10 spline basis functions we use 15, as this provides slightly better results along with superior computational stability. We note that the approach of   \cite{everitt2013timefrequency} is not considered here due to the fact that the parameterization of the spectral density through the Wittle likelihood requires subjective knowledge.

The second approach is the AutoPARM method \citep{davis2006break}. Although this approach combines the GA and MDL to automatically search for potential break points along with the AR orders for each segment, four parameters are crucial for the GA: the number of islands, the number of chromosomes in each island, the number of generations for migration, and the number of chromosomes replaced in a migration; see \citet{davis2006break} for a comprehensive discussion. All of these parameters were chosen identical to those used in  \cite{davis2006break}. 

The third approach is the AutoSLEX method \citep{ombao2001auto}. Given a fixed value for the complexity penalty parameter of the cost function, AutoSLEX can automatically segment a given signal and choose a smoothing parameter. Following the suggestion of \cite{ombao2001auto}, we set this parameter equal to one.
The last method we consider is the approach of \cite{west1999evaluation}, referred to as WPK1999. This approach requires specification of three parameters: the TVAR order and two discount factors --  one associated with the variance of the time-varying coefficients and the other with the innovation variances. In general, the discount factor values are in the range $0.9-0.999$ \citep{west1999evaluation}. Therefore, for our simulations, we give each discount factor a set of values from $0.8$ to $1$ (with equal spacing of $0.02$) and, further, a set of values for the TVAR order from $1$ to $15$. Given these values, we choose the combination that achieves the maximum likelihood \citep{west1999evaluation}.         

Our approach uses the two selection methods discussed in Section~\ref{selection} to search for the TVAR order with appropriate discount factor values. The selected combination of $(P,\gamma,\delta)$ with  $\gamma$ and $\delta$ held fixed over all stages of the Bayesian lattice filter is referred to as BLFFix. The selected combination of ($P,\gamma_{m},\delta_{m}$), for $m=1,\ldots,P$, is referred to as BLFDyn. Again, the candidate space of parameters for both discount factors are from $0.8$ to $1$ (with equal spacing of $0.02$), along with orders from $1$ to $15$.      

We consider four types of nonstationary signals: 1) TVAR of order $2$ with constant innovation variance; 2) TVAR of order $6$ with constant innovation variance; 3) a piecewise AR process with constant innovation variance; and 4) simulated signals based on an {\it{Enchenopa}} treehopper communication signal \citep{holan2010modeling}; see Section~\ref{sec:bugs}. Each simulation consists of $200$ realizations. To evaluate the performance in estimating the various time-frequency representations, we calculate the average squared error (ASE) for each realization as follows \citep{ombao2001auto}  
\begin{equation}
\mbox{ASE}_{n}= \left ( TL \right )^{-1}\sum_{t=1}^{T}\sum_{l=1}^{L}\left(\mbox{log}\widehat{S}(t,\omega_{l}) - \mbox{log}S(t,\omega_{l}) \right)^{2},
\label{eq:ASEn}
\end{equation}
where $n=1,\ldots,200$, $\omega_{l}=0,0.005,\ldots,0.5$, and $T$ denotes the length of the simulated series.  Lastly, we denote $\overline{\mbox{ASE}}=(1/200)\sum_{n=1}^{200} \mbox{ASE}_{n}$.  For each simulation study, Table~\ref{tab:meanASE} summarizes the mean values and standard deviations for $\mbox{ASE}_{n}$.

In the case of the AutoSLEX method the number of frequencies in (\ref{eq:ASEn}) differs from the other approaches considered.  In particular, the AutoSLEX approach dyadically segments the signal up to a given maximum scale $J$ such that $2^J$ is less than signal length, $T$. Subsequently, AutoSLEX automatically determines whether a segment at a particular scale will be included in final segmentation. Once this has been completed, the frequency resolution for the AutoSLEX approach is equal to $T/2^{(j+1)}$ where $j$ is the scale of the largest segment included in the final segmentation.

\subsection{Time-Varying AR(2) Process}\label{subsec:tvar2sim}
\begin{figure}[h] 
\centerline{
\includegraphics[width=0.8\textwidth, angle = -90]{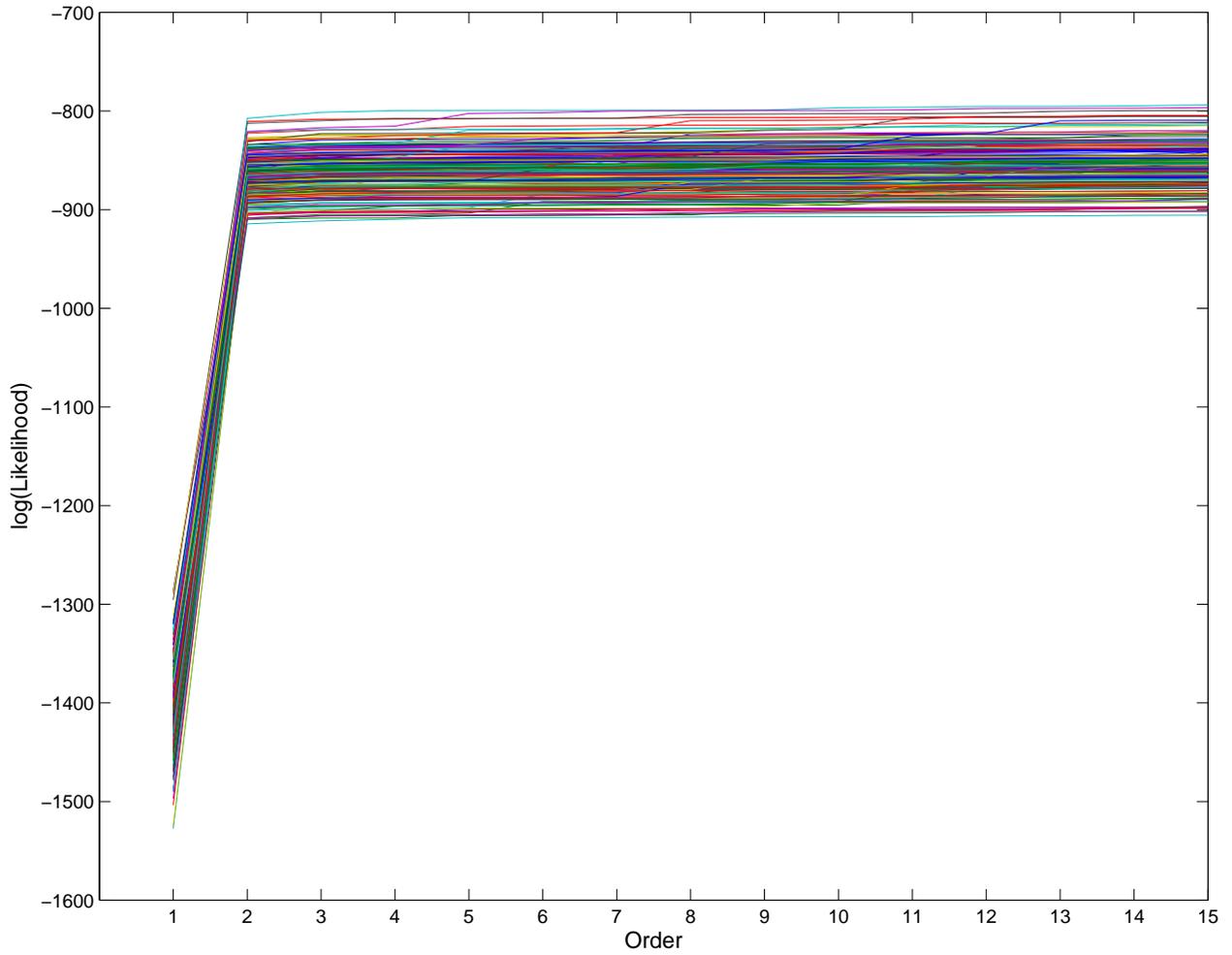}}
\caption{BLF-scree plot of the 200 realizations of the time-varying AR(2) process (TVAR2) given in Section~\ref{subsec:tvar2sim}.  \baselineskip=10pt } \label{fig:screeTVAR2}
\end{figure}

\begin{figure}
\centerline{
\includegraphics[width=0.8\textwidth,angle=-90]{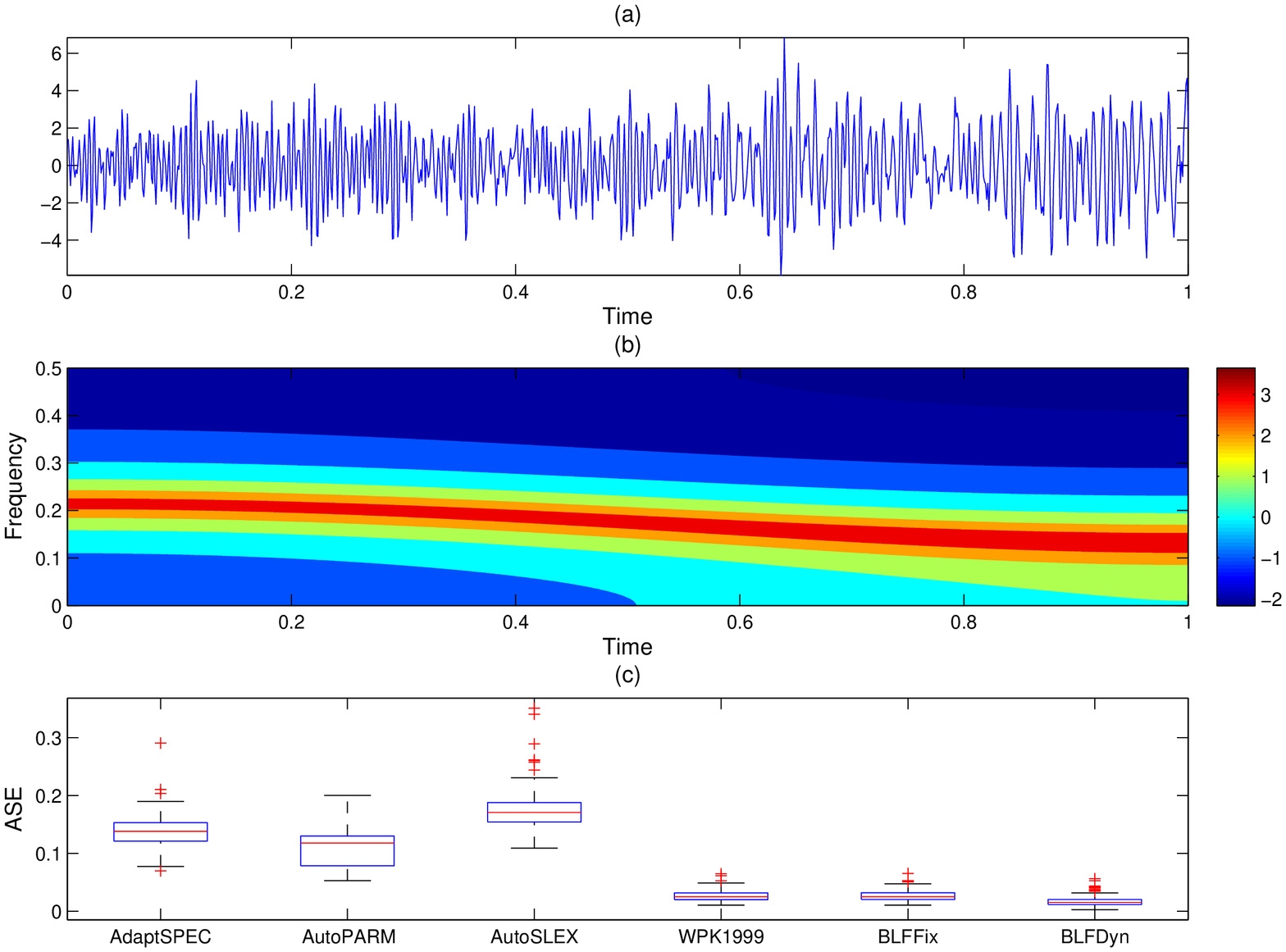}}
\caption{(a) and (b) depict one realization along with the true time-frequency representation of the time-varying AR(2) process (TVAR2), respectively (Section~\ref{subsec:tvar2sim}). (c) illustrates the box-plots of the average squared error (ASE) values corresponding to the time-frequency representation of the TVAR2 for all of the approaches considered.  \baselineskip=10pt } \label{fig:tvar2}
\end{figure}

We simulate signals from the same time-varying AR(2) process (TVAR2), used in \cite{davis2006break} and \cite{rosen2009localspectral, rosen2012adaptspec}, which is defined as follows
\begin{align*}
x_{t} & = a_{t}x_{t-1} - 0.81x_{t-2}+\epsilon_{t},\\
a_{t} & = 0.8(1-0.5\mbox{cos}(\pi t/1024)),
\end{align*}
where $\epsilon_{t}\iid N(0,1)$ and $t=1,\ldots,1024$. Figure~\ref{fig:screeTVAR2} shows the BLF-scree plot, suggesting that order two is the appropriate choice for all 200 realizations. Since the time-varying coefficient $a_{t}$, varies slowly with time, this process naturally exhibits a slowly evolving time-varying spectrum (Figure~\ref{fig:tvar2}). The box-plots of the ASE values in Figure~\ref{fig:tvar2} show that the group of TVAR-based models (i.e., WPK1999, BLFFix, and BLFDyn) perform superior to the group of non-TVAR-based models (i.e., AdaptSPEC, AutoPARM, and AutoSLEX), with BLFDyn performing the best. In particular, there is a significant percent reduction in $\overline{\mbox{ASE}}$ for BLFDyn relative to the other methods considered (Table~\ref{tab:meanASE}).

\subsection{Time-Varying AR(6) Process}\label{subsec:tvar6sim}
We consider signals from the same time-varying AR(6) process of order six (TVAR6) used in \cite{rosen2009localspectral}. This time-varying AR(6) process can be compactly expressed as $\phi_{t}(B)x_{t}=\epsilon_{t}$, $t=1,\ldots,T$, in terms of a characteristic polynomial function $\phi_{t}(B)$, with $\epsilon_{t}\iid N(0,1)$ and $B$ the backshift operator (i.e., $B^{p}x_{t}=x_{t-p}$). The characteristic polynomial function for this process can be factorized as
\begin{equation*}
\phi_{t}(B) = (1-a_{t,1}B)(1-a_{t,1}^{*}B)(1-a_{t,2}B)(1-a_{t,2}^{*}B)(1-a_{t,3}B)(1-a_{t,3}^{*}B),
\end{equation*}
where the superscript $*$ denotes the complex conjugate. Also, for $p=1,2,3$, let $a_{t,p}^{-1}= A_{p}\mbox{exp}(2\pi i\theta_{t,p})$, where the $\theta_{t,p}$s are defined by
\begin{align*}
\theta_{t,1} & = 0.05 + (0.1/(T-1))t,\\
\theta_{t,2} & = 0.25,\\
\theta_{t,3} & = 0.45 - (0.1/(T-1))t,
\end{align*}
with $T=1024$ and the values of $A_{1}$, $A_{2}$, and $A_{3}$ equal to $1.1$, $1.12$, and $1.1$, respectively.  

\begin{figure}
\centerline{
\includegraphics[width=0.8\textwidth,angle=-90]{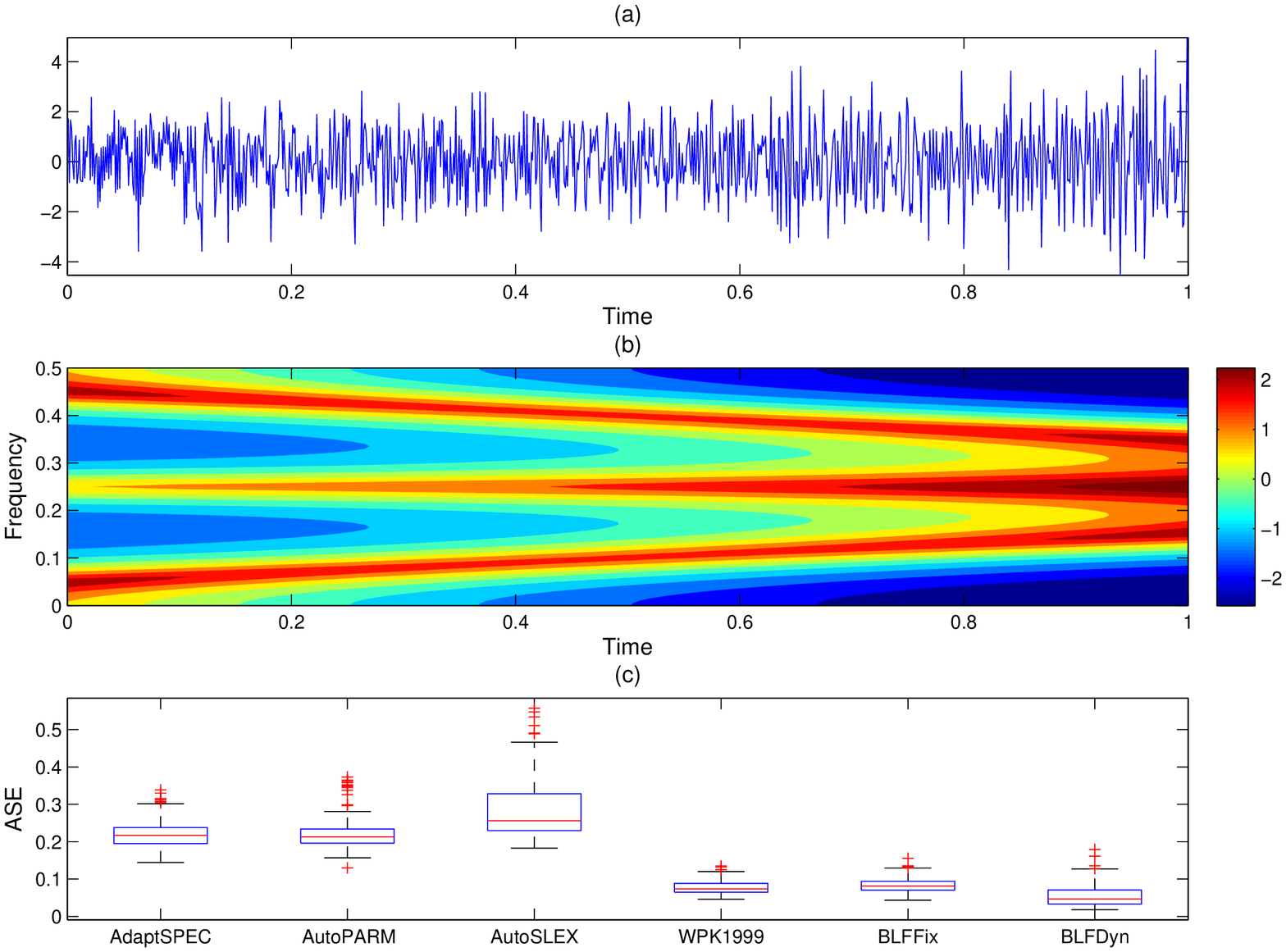}}
\caption{(a) and (b) depicts one realization along with the true time-frequency representation of the time-varying AR(6) process (TVAR6), respectively (Section~\ref{subsec:tvar6sim}). (c) illustrates the box-plots of the average squared error (ASE) values corresponding to the time-frequency representation of the TVAR6 for all of the approaches considered.  \baselineskip=10pt } \label{fig:tvar6}
\end{figure}

The BLF-scree plot (not shown) suggests that order six is the appropriate choice for all 200 realizations.
The TVAR(6) contains three pairs of time-varying conjugate complex roots.  Figure~\ref{fig:tvar6} illustrates that the TVAR(6) has a time-varying spectrum with three peaks. Similar to the TVAR(2) analysis (Section~\ref{subsec:tvar2sim}), the TVAR-based models outperform the group of non-TVAR-based models, with BLFDyn performing superior to the others.  Again, the percent reduction in $\overline{\mbox{ASE}}$ is 
significant relative to the other approaches considered (Table~\ref{tab:meanASE}).
       
\subsection{Piecewise Stationary AR Process}\label{subsec:pwarsim} 
\begin{figure}
\centerline{
\includegraphics[width=0.8\textwidth,angle=-90]{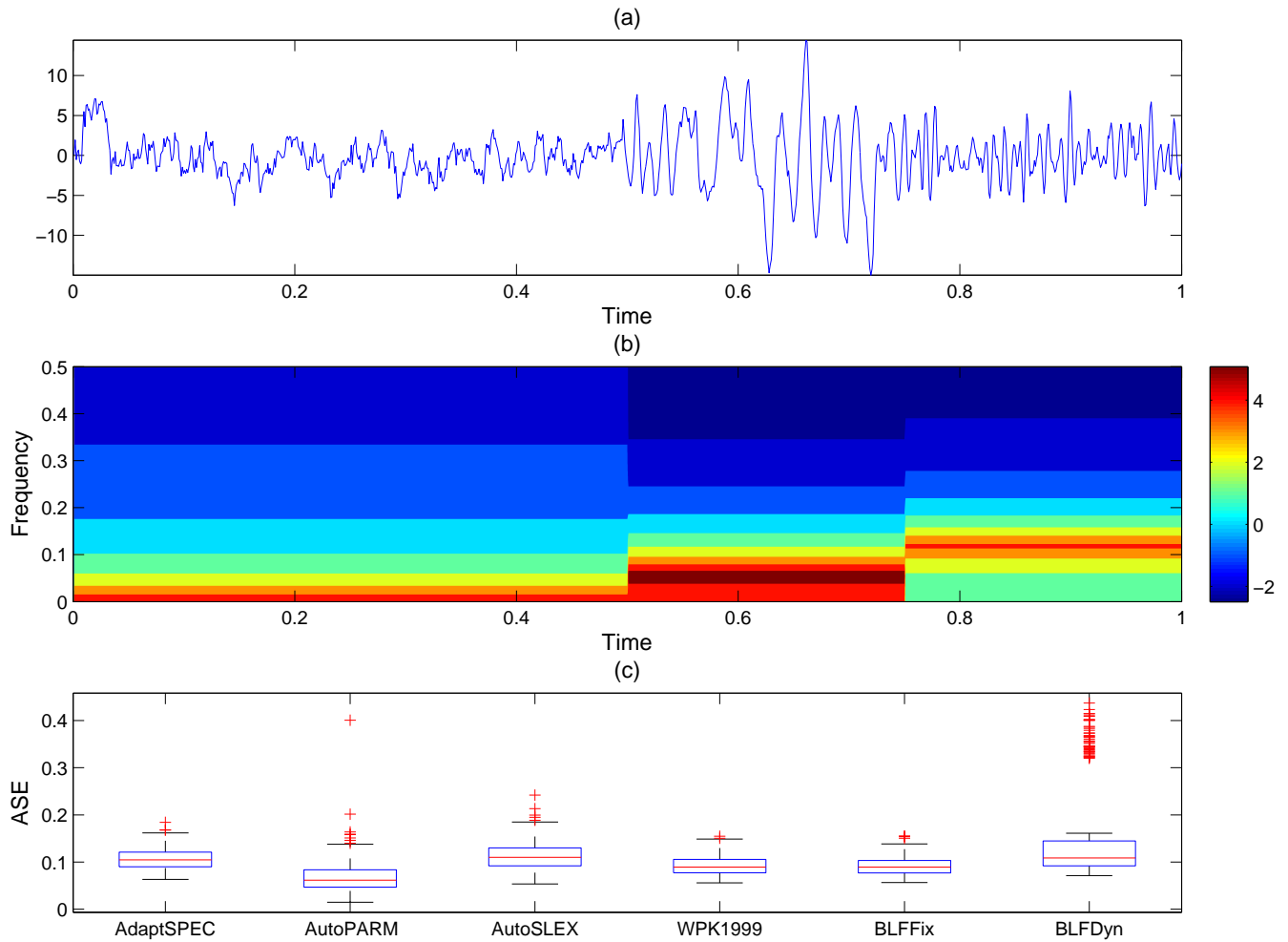}}
\caption{(a) and (b) depict one realization along with the true time-frequency representation of the piecewise AR process (PieceAR), respectively (Section~\ref{subsec:pwarsim}). (c) illustrates the box-plots of the average squared error (ASE) values corresponding to the time-frequency representation of the PieceAR for all of the approaches considered.  \baselineskip=10pt } \label{fig:piece}
\end{figure}

The signals simulated here are based on the same piecewise stationary AR process, used by \cite{davis2006break} and \cite{rosen2009localspectral, rosen2012adaptspec} and is defined as follows 
\begin{eqnarray*} 
x_{t}= \left\{ \begin{array}{rll} 
                         0.9x_{t-1}+\epsilon_{t}; & \mbox{if} & 1 \leq t \leq 512, \\
                         1.69x_{t-1}-0.81x_{t-2}+\epsilon_{t}; & \mbox{if} & 513 \leq t \leq 768, \\
1.32x_{t-1}-0.81x_{t-2}+\epsilon_{t}; & \mbox{if} & 769 \leq t \leq 1024,  
                  \end{array}\right. 
\end{eqnarray*}
where $\epsilon_{t}\iid N(0,1)$. These generated signals are referred to as PieceAR. Since it is difficult to choose the order for some realizations visually using the BLF-scree plot, we use \eqref{eq:criteria}, with $\tau=0.5$,  to choose the order. The numerical method suggests order two for some realizations and order three for the others. The true process includes three segments, with each of the segments mutually independent. The piecewise nature of this process is clearly depicted by its time-varying spectrum (Figure~\ref{fig:piece}b). The box-plot (Figure~\ref{fig:piece}c) shows that AutoPARM exhibits superior performance in terms of the smallest median ASE. However, WPK1999 and BLFFix may perform more robustly (i.e., less outlying ASE values).  Although we see a 23.78\% reduction in $\overline{\mbox{ASE}}$ for AutoPARM versus BLFFix, we find that BLFFix performs superior to the remainder of the approaches and has the smallest standard deviation across the 200 simulations.  Table~\ref{tab:meanASE} summarizes the mean values and standard deviations for this simulation. The findings here are not surprising as AutoPARM is ideally suited toward identifying and estimating piecewise AR processes.  For our approach, taking $(\gamma, \delta)$ fixed is advantageous for processes that are not slowly-varying.

\subsection{Simulated Insect Communication Signals}\label{subsec:bugsim}  
\begin{figure}
\centerline{
\includegraphics[width=0.8\textwidth,angle=-90]{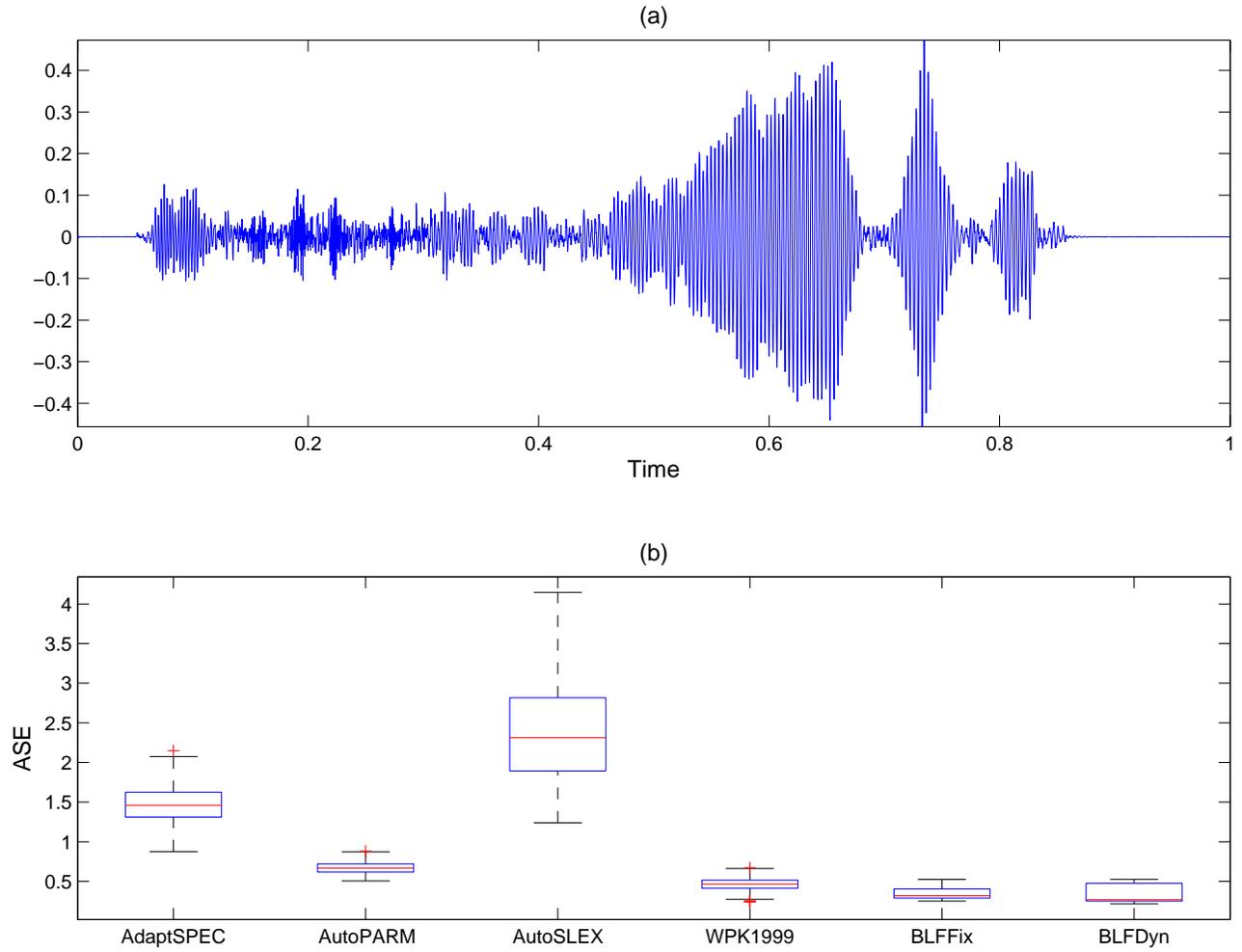}}
\caption{(a) depicts one realization of the simulated insect communication signals (SimBugs), Section~\ref{subsec:bugsim}. (b) illustrates the box-plots of the average squared error (ASE) values corresponding to the time-frequency representation of the SimBugs for all of the approaches considered.  \baselineskip=10pt } \label{fig:simBugs} 
\end{figure}

The signals considered in this simulation are formulated such that they exhibit the same properties as an {\it{Enchenopa}} treehopper mating signal; see Section~\ref{sec:bugs} for a complete discussion.  Specifically, we fit a TVAR(6) model to the signal $x_t$, $t=1,\ldots, 4096$, to obtain time-varying AR coefficients and innovation variances.  Typically, with these type of nonstationary signals, the innovation variances are time dependent, which is markedly different from the previous examples where the innovation variance was constant. The signals generated by these parameters are referred to as SimBugs. As expected, the BLF-scree plot (not shown) suggests that order six may be an appropriate choice for all 200 realizations. Figure~\ref{fig:simBugs}a illustrates one realization of the SimBugs, whereas Figure~\ref{fig:simBugs}b provides box-plots that characterize the distribution of $\mbox{ASE}_n$ over the 200 simulated signals. Specifically, from  Figure~\ref{fig:simBugs}b, we see that BLFFix and BLFDyn perform better than the other approaches, in terms of median $\mbox{ASE}$. Further, we find that BLFFix and BLFDyn are similar, in terms of $\overline{\mbox{ASE}}$, although the median of BLFDyn (0.2675) is smaller than that of BLFFix (0.3176). 

Table~\ref{tab:meanASE} summarizes the mean values and standard deviations for this simulation.  From this table we see that the reduction in $\overline{\mbox{ASE}}$ for BLFDyn is 6.01\% over BLFFix and that both exhibit a substantial percent reduction in $\overline{\mbox{ASE}}$ over the other methods considered.  

\section{Case Studies}\label{sec:application}
\subsection{Animal Communication Signals}\label{sec:bugs}
\begin{figure}
\centerline{
\includegraphics[width=0.8\textwidth,angle=-90]{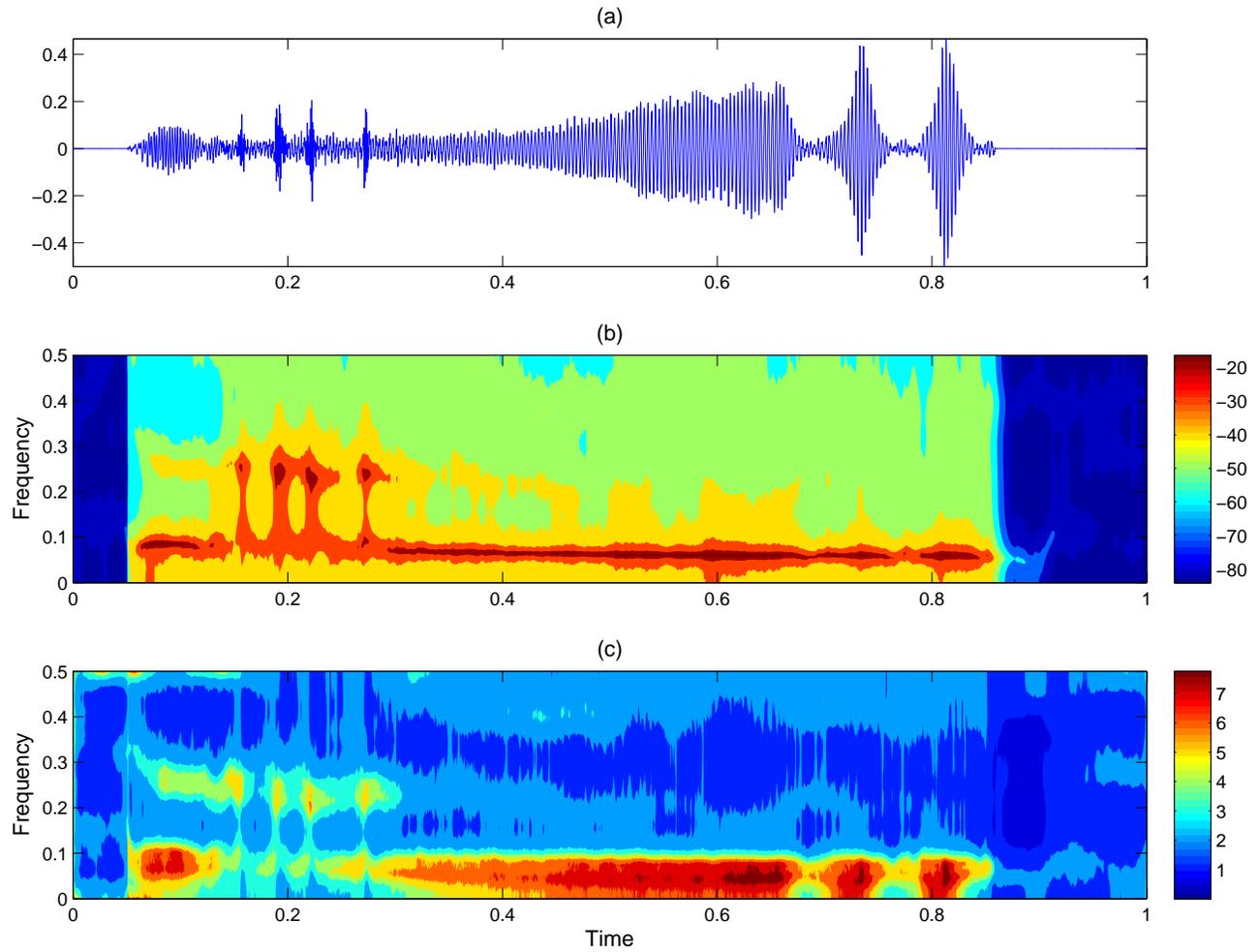}}
\caption{(a) An example of typical signal corresponding to a successful mater (Section~\ref{sec:bugs}). (b) and (c) present posterior mean and standard deviation of the TVAR(6) spectral representation of the signal in plot (a). \baselineskip=10pt } \label{fig:rawtree}
\end{figure}

Understanding the dynamics of populations is an important component of evolutionary biology. Many organisms exhibit complex characteristics that intricately relate to fitness. For example, the mating signal of the {\it{Enchenopa}} treehopper represents a phenotype of the insect that is used in mate selection. During the mating season, males in competition deliver their vibrational signals through stems of plants to females \citep[see][ and the references therein]{cocroft2006vib}. The data considered here comes from an experiment that was previously analyzed in \citet{cocroft2006vib} and \citet{holan2010modeling}. The experiment was designed with the goal of reducing potential confounding effects between environmental and phenotypical variation. In this experiment, males signals were recorded one week prior to the start of mating. Figure~\ref{fig:rawtree}a displays a typical signal of from a successful mater, with length 4,739 downsampled from registered signals of length 37,912.   Justification for the appropriateness of downsampling the original signal, in this context, can be found in \cite{holan2010modeling}. Also, as discussed in \cite{holan2010modeling}, this signal shows a series of broadband clicks preceding a frequency-modulated sinusoidal component, followed by a series of pulses.

\begin{figure}
\centerline{
\includegraphics[width=0.8\textwidth,angle=-90]{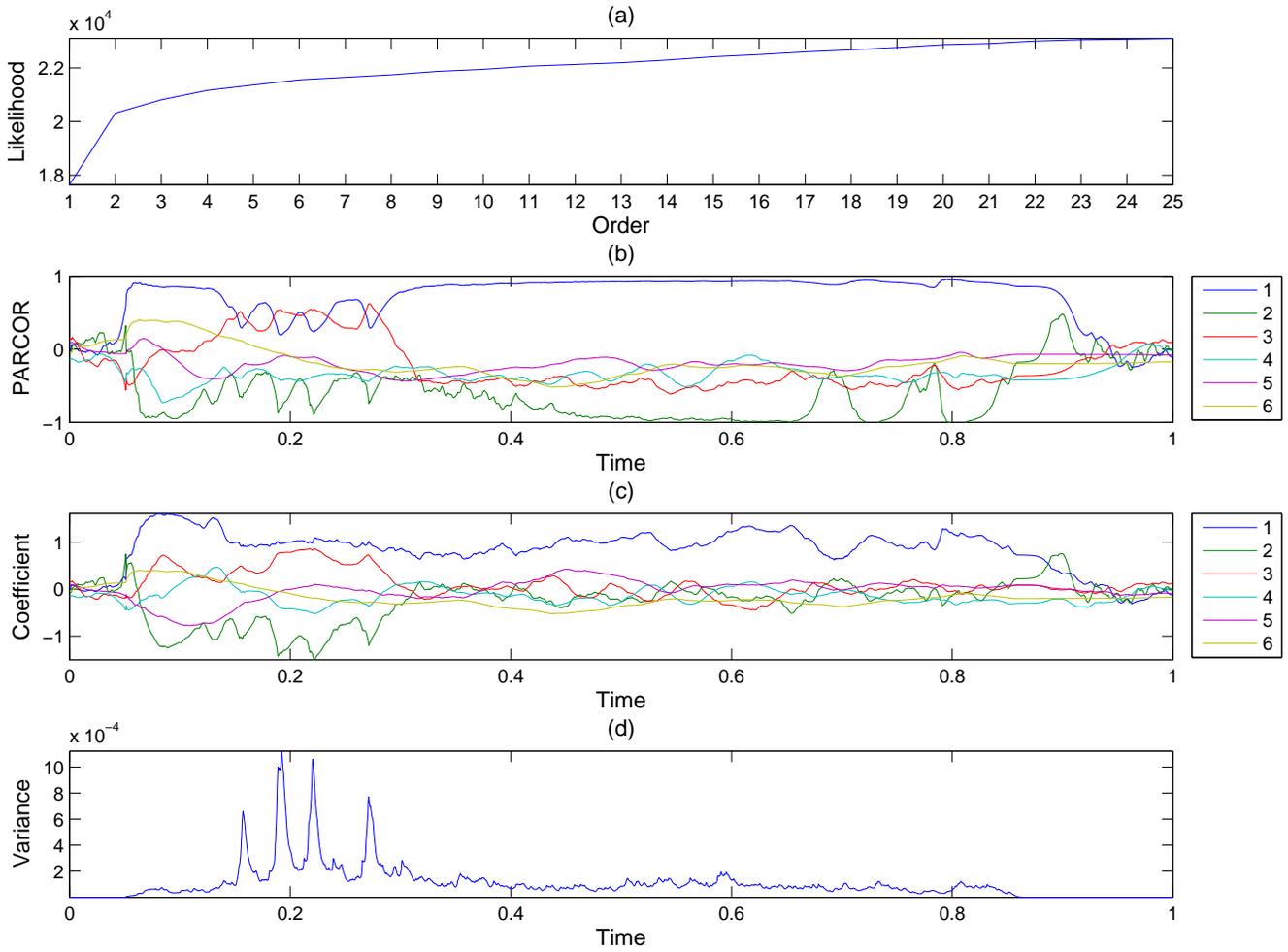}}
\caption{(a) shows the BLF-scree plot of the treehopper communication signal. (b) depicts the first six time-varying estimated PARCOR coefficients. (c) and (d) show the estimated time-varying coefficients and innovation variances of the TVAR(6) model.  \baselineskip=10pt } \label{fig:coefsNvar}
\end{figure}

For this analysis, we used the BLFDyn approach to search for a model having both discount factors in the range of $0.8$ to $1$ (with equal spacing of $0.02$) and an order between $1$ and $25$. Figure~\ref{fig:coefsNvar}a  shows an increase in $\Lagr_{m}$ along with the order, which is different from the simulated TVAR(6) model in Section~\ref{subsec:bugsim}. Hence, we use \eqref{eq:criteria} with the model order chosen by $\tau=0.5$. This rule yields a TVAR(6) model. In  Figure~\ref{fig:coefsNvar}b, the PARCOR coefficients of lag larger than two are close to zero following time around $0.3$ (where the time axis has been normalized such that $t\in(0,1)$). Thus, the last four TVAR coefficients after time $0.3$ are close to zero (Figure~\ref{fig:coefsNvar}c). Such phenomena suggests that the period before time $0.3$ has a more complex dependence structure. Figure~\ref{fig:coefsNvar}d illustrates that the innovation variance exhibits higher volatility at the beginning signal. These bursts in the innovation variance are related to the series of broadband clicks at the beginning of the signal. Finally, Figure~\ref{fig:rawtree}b presents the time-frequency representation of the treehopper signal using the TVAR(6) model and corroborates the significance of the broadband clicks at the beginning of the signal. Figure~\ref{fig:rawtree}c shows the posterior standard deviation of the time-frequency representation using 2,000 MCMC samples from 210,000 MCMC iterations by discarding the first 10,000 iterations as burn-in and keeping every 100-th iteration of the remainder.        

\subsection{Wind Components}\label{sec:winds}
\begin{figure}
\centerline{
\includegraphics[width=0.8\textwidth,angle=-90]{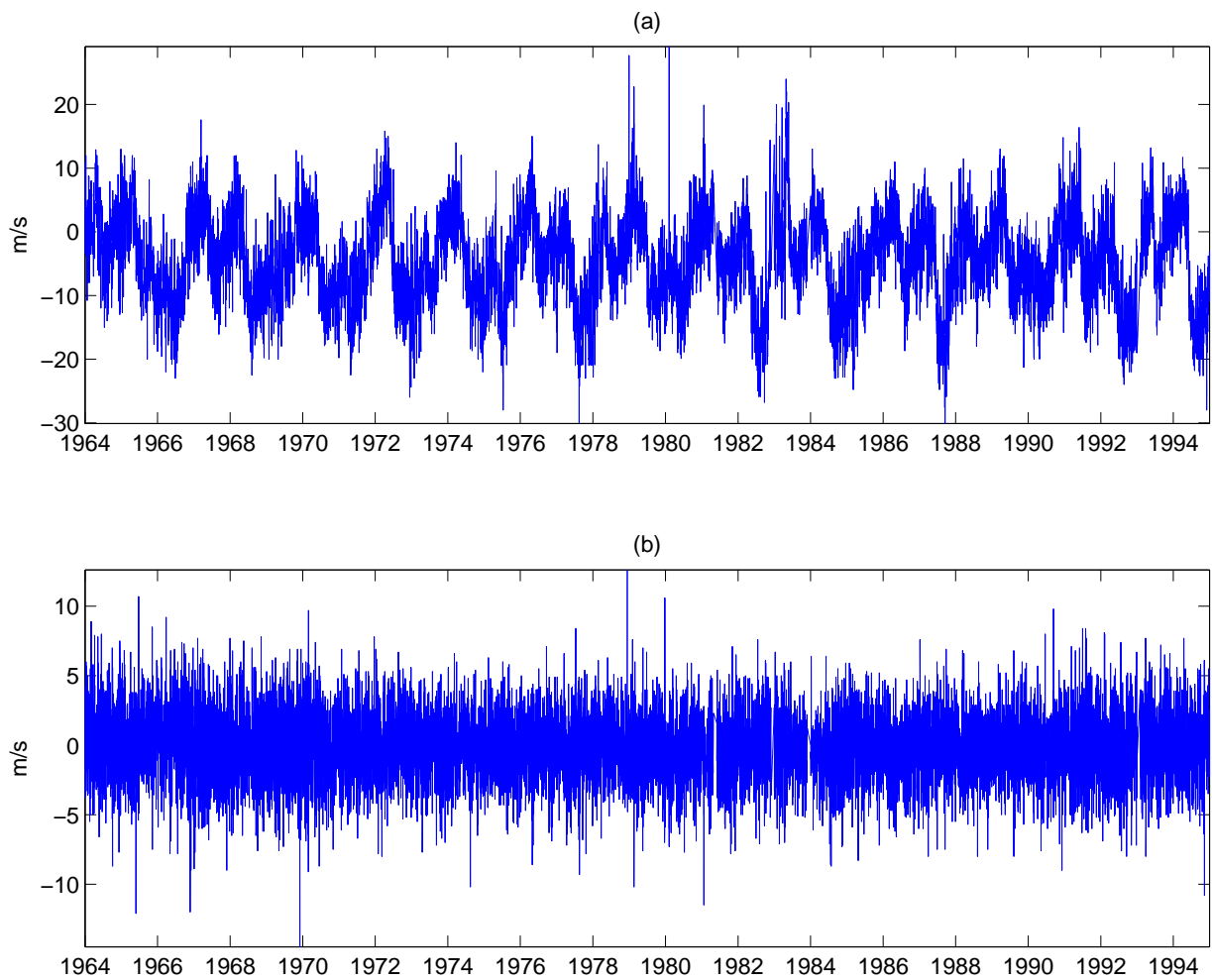}}
\caption{\baselineskip=10pt (a) and (b) show daily time series (1964-1994) of east/west and north/south components of wind, respectively.  Both components are measured in meters per second (m/s). } \label{fig:rawWind}
\end{figure}

We study the time-frequency representations of the east/west and north/south wind components, recorded daily at Chuuk Island in the tropical Pacific during the period of 1964 to 1994 \citep[see][Sections 3.5.3 and 3.5.4]{cressiewikle2011}. The data studied are at the level of 70 hPa, which is important scientifically due to the likely presence of westward and eastward propagating tropical waves, and the presence of the quasi-biennial oscillation (QBO) \citep{wikle1997wind}. Figure~\ref{fig:rawWind} shows the two wind component time series from which we can discern visually that the east/west series clearly exhibits the QBO signal, but no discernible smaller-scale oscillations are present in either series. Our interest is then whether the time-varying spectra for these series suggest the presence of time-varying oscillations, which are theorized to be present.

We consider the same search space for the model order and discount factors as that used for the treehopper communication signal (Section~\ref{sec:bugs}).  The BLF-scree plot for the east/west component shows an increase of $\Lagr_{m}$ along with the order. Therefore, we use \eqref{eq:criteria} with $\tau=0.5$ and choose the order equal to three. The PARCOR coefficient of lag one depends on time but the PARCOR coefficient of lag two and three appear to be constant. The innovation variances of TVAR(3) model are time dependent. On the other hand, the BLF-scree plot for the north/south component shows a turning point at order four so that we choose the order equal to four. The PARCOR coefficients are time independent but the innovation variances of the TVAR(4) model are time dependent. The preceding findings are illustrated in Figures~\ref{fig:ewWind} and \ref{fig:nsWind} of the Appendix.  Figures~\ref{fig:Windspec}a and \ref{fig:Windspec}b show the estimated time-frequency representations of the wind east/west and north/south components. Figures~\ref{fig:Windspec}c and \ref{fig:Windspec}d present the associated posterior standard deviations of the time-frequency representation using 2,000 MCMC samplers from 210,000 MCMC iterations by discarding the first 10,000 iterations as burn-in and keeping every 100-th iteration of the remainder. The east/west component time-varying spectrum does suggest that the QBO intensity varies considerably as evidenced by the power in the low-frequencies. Perhaps more interesting is the suggestion of time-varying equatorial waves in the north/south wind component time-varying spectrum. In particular, the lower-frequency (Kelvin and Rossby) waves with frequencies between 0.1 and 0.2 show considerable variation in duration of wave activity, as well as intensity. One also sees time-variation in the likely mixed-Rossby gravity waves in the frequency band between 0.2 and 0.35. Interestingly, in some cases these are in phase with the lower-frequency wave activity but more often act in opposition. We also note the almost complete collapse of the equatorial wave activity centered on 1984.    

\subsection{Economic Index}\label{sec:ecod}
\begin{figure}
\centerline{
\includegraphics[width=0.8\textwidth,angle=-90]{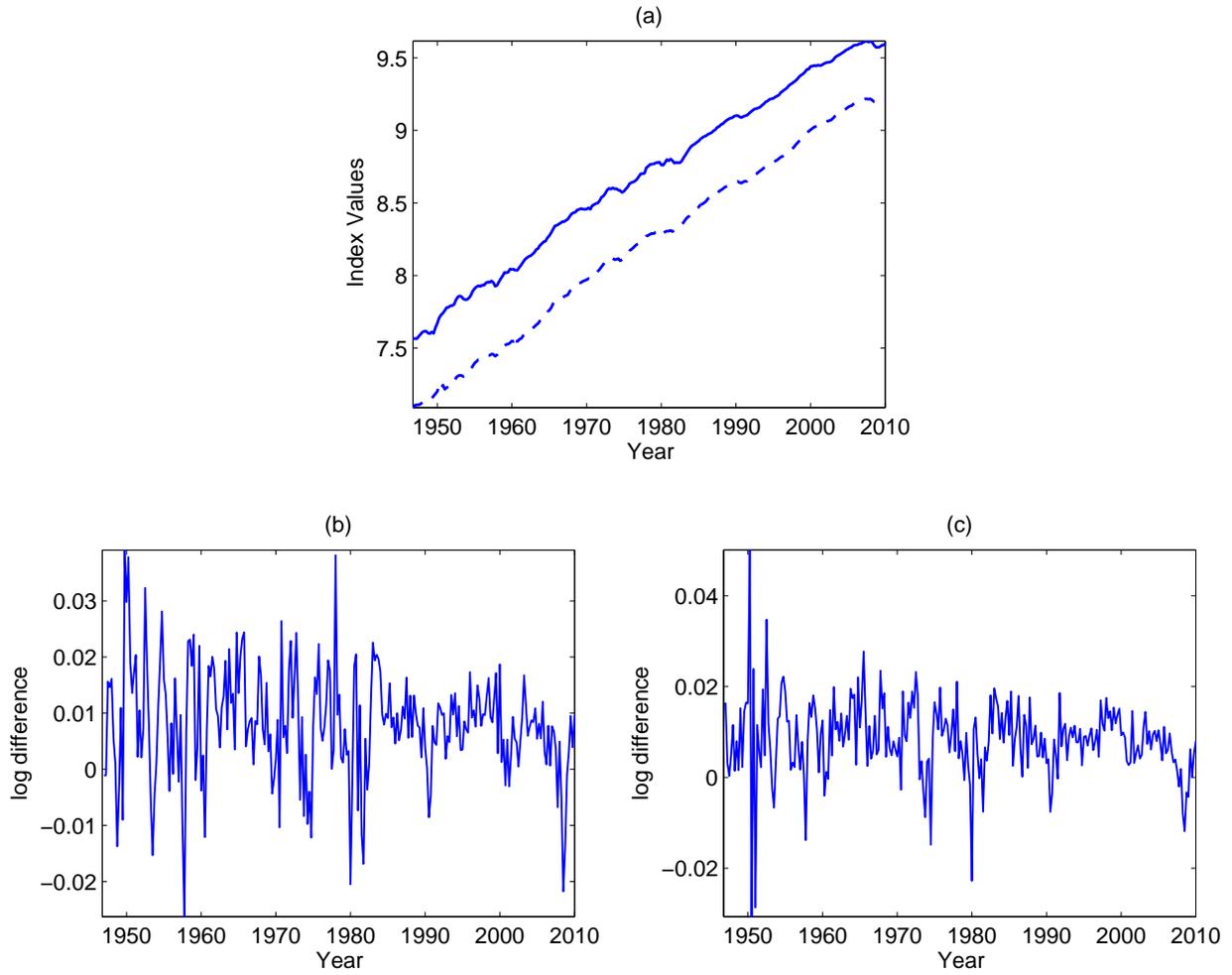}}
\caption{(a) displays the logarithm of GDP (solid line) and consumption (dash line) in from the first quarter of 1947 to the first quarter of 2010. (b) and (c) present the log difference of GDP and the log difference of consumption. \baselineskip=10pt } \label{fig:loggdpcon}
\end{figure}

\cite{koopman2011kalman} study the time-frequency representation of the log difference of the US gross domestic product (GDP), consumption, and investment from the first quarter of 1947 to the first quarter of 2010. For comparison, we obtained the series from the Federal Reserve Bank of St. Louis \href{http://research.stlouisfed.org/fred2}{$($http://research.stlouisfed.org/fred2$)$}. Since the investment series is unavailable from the website, we only consider the GDP and consumption in our study. Figure~ \ref{fig:loggdpcon}a illustrates the trend of the logarithm of the GDP and consumption series. Similar to \cite{koopman2011kalman}, we analyzed the log difference of these two series (Figures~\ref{fig:loggdpcon}b and \ref{fig:loggdpcon}c).

\begin{figure}
\centerline{
\includegraphics[width=0.8\textwidth,angle=-90]{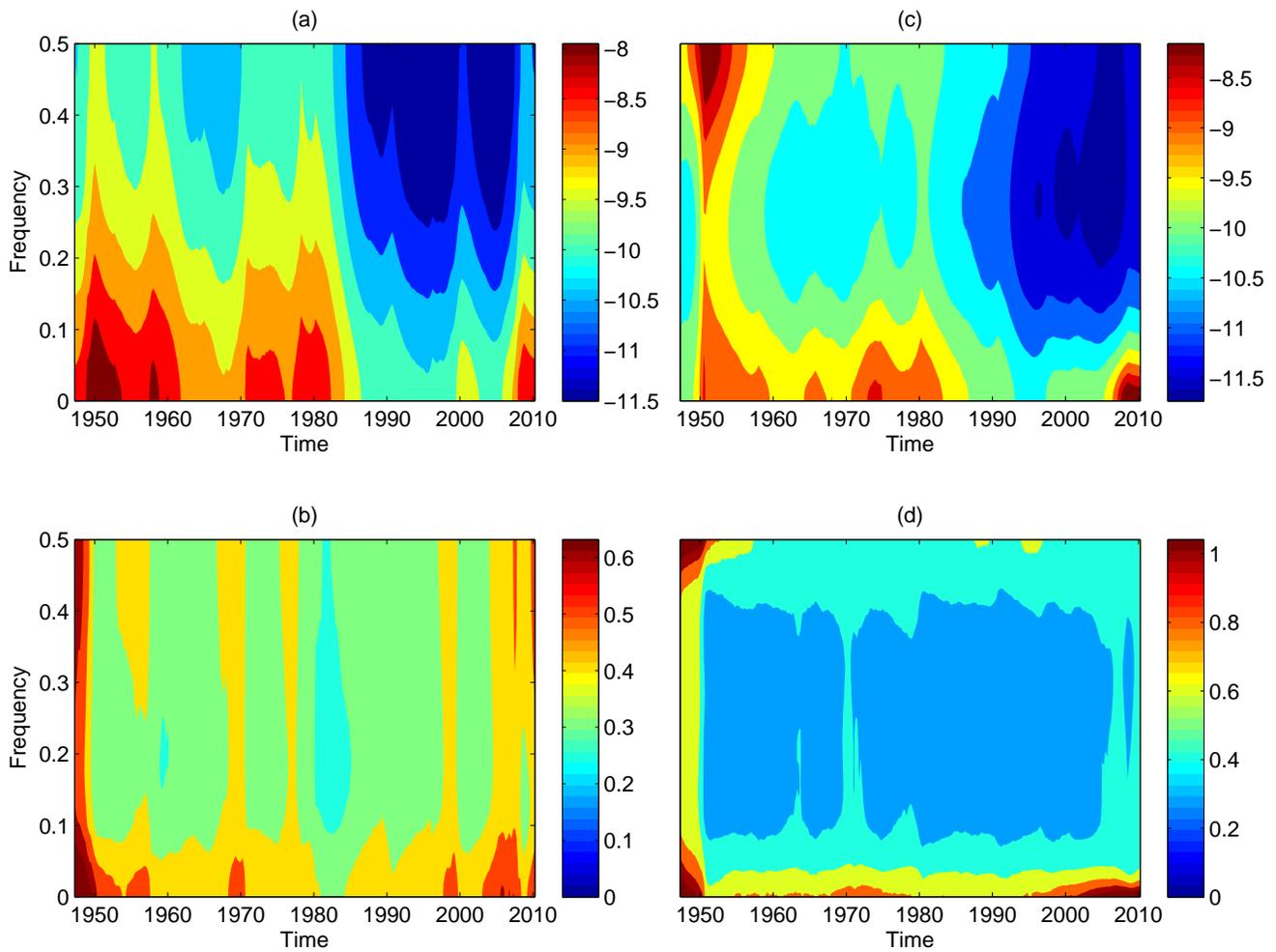}}
\caption{(a) and (b) display the time-frequency representations of the log difference of GDP and consumption series, respectively. (c) and (d) show the standard deviations of the time-frequency representations of the log difference of GDP and consumption series, respectively.   \baselineskip=10pt } \label{fig:gdp_consump}
\end{figure}

We consider the same search space for the model order and discount factors as that used for the treehopper communication signal (Section~\ref{sec:bugs}). By using \eqref{eq:criteria} with $\tau=0.5$, we choose TVAR(1) for the log difference of GDP and TVAR(2) for the log difference of consumption. Figures~\ref{fig:gdp_consump}a and \ref{fig:gdp_consump}b show the estimated time-frequency representations and the associated standard deviations of the log difference of GDP and consumption series, respectively. Similar to the results of \cite{koopman2011kalman}, both the US macroeconomic indices exhibit relatively larger spectra (or fluctuations) for the early period around 1950. Then, due to the oil crisis, another fluctuation comes out in the period of 1970--1980. The fluctuation at the end of the sample reflects the 2008--2009 worldwide financial crisis. See \cite{koopman2011kalman} for further discussion.  Using the same MCMC iterations as Section~\ref{sec:bugs}, Figures~\ref{fig:gdp_consump}c and \ref{fig:gdp_consump}d present the associated posterior standard deviations.

\section{Discussion}\label{sec:discussion}
This paper develops a computationally efficient method for model-based time-frequency analysis. Specifically, we consider a fully Bayesian lattice filter approach to estimating time-varying autoregressions. By taking advantage of the partial autocorrelation domain, our approach is extremely stable.  That is, the PARCOR coefficients and the TVAR innovation variances are specified within the lattice structure and then estimated simultaneously.  Notably, the full conditional distributions arising from our approach are all of standard form and, thus, facilitate easy estimation.

The framework we propose extends the current model-based approaches to time-frequency analysis and, in most cases, provides superior performance, as measured by the average squared error between the true and estimated time-varying spectral density. In fact, for slowly-varying processes we have demonstrated significant estimation improvements from using our approach. In contrast, when the true process comes from a piecewise AR model the approach of \citet{davis2006break} performed best, with our approach a close competitor and performing second best. This is not unexpected as the AutoPARM method is a model-based segmented approach and more closely mimics the behavior of a  piecewise AR.

In addition to a comprehensive simulation study we have provided three real-data examples, one from animal (insect) communication, one from environmental science, and one from macro-economics.  In all cases, the exceptional time-frequency resolution obtained using our approach helps identify salient features in the time-frequency surface. Finally, as a by-product of taking a fully Bayesian approach, we are naturally able to quantify uncertainty and, thus, use our approach to draw inference.

\section*{Acknowledgments}
This research was partially supported by the U.S. National Science Foundation (NSF) and the U.S. Census Bureau under NSF grant SES-1132031, funded through the NSF-Census Research Network (NCRN) program.  The authors would like to thank the Reginald Cocroft lab for use of the insect communication data analyzed in Section~\ref{sec:bugs}. The authors would like to thank Drs. Richard Davis, Hernando C. Ombao, and Ori Rosen for providing their codes.  Finally, we also thank Dr. Mike West for generously providing his code online.

\vskip .35in
\begin{table}[htdp] 
\caption{\baselineskip=10pt The mean, $\overline{\mbox{ASE}}$ and standard deviation, $\mbox{sd}_{ASE}$, of the ASE values for the simulations presented in Section~\ref{sec:simulation}. Note that the bold values represent the approach having minimum $\overline{\mbox{ASE}}$. }
\begin{center}
\begin{scriptsize}
\begin{tabular}{ c| c c c c c c}
\hline
 & \multicolumn{6}{c}{$\overline{\mbox{ASE}}$($\mbox{sd}_{ASE}$)}\\ 
 & AdaptSPEC & AutoPARM & AutoSLEX & WPK1999 & BLFFix & BLFDyn\\ 
\hline
 TVAR2 & 0.1383 (0.0275) & 0.1085 (0.0302)&0.1735 (0.0333) & 0.0268 (0.0093) & 0.0269 (0.0093) & {\bf 0.0170 (0.0084)}\\
 TVAR6 & 0.2195 (0.0351) & 0.2233 (0.0439)&0.2885 (0.0796) & 0.0771 (0.0171) & 0.0841 (0.0191) & {\bf 0.0543 (0.0276)}\\
 PieceAR & 0.1070 (0.0227) & {\bf 0.0702 (0.0392)}& 0.1141 (0.0296) & 0.0931 (0.0218) & 0.0921 (0.0205) & 0.1607 (0.1107)\\
 SimBugs & 1.4760 (0.2237) & 0.6711 (0.0725)& 2.3993 (0.6078) & 0.4627 (0.0796) & 0.3444 (0.0659) & {\bf 0.3237 (0.1075)}\\
\hline
\end{tabular}
\end{scriptsize}
\end{center}
\label{tab:meanASE}
\end{table}

\clearpage\pagebreak\newpage 

\clearpage\pagebreak\newpage 

\clearpage\pagebreak\newpage 

\clearpage\pagebreak\newpage 

\clearpage\pagebreak\newpage 

\clearpage\pagebreak\newpage 

\clearpage\pagebreak\newpage 

\clearpage\pagebreak\newpage 
\begin{figure}
\centerline{
\includegraphics[width=0.8\textwidth,angle=-90]{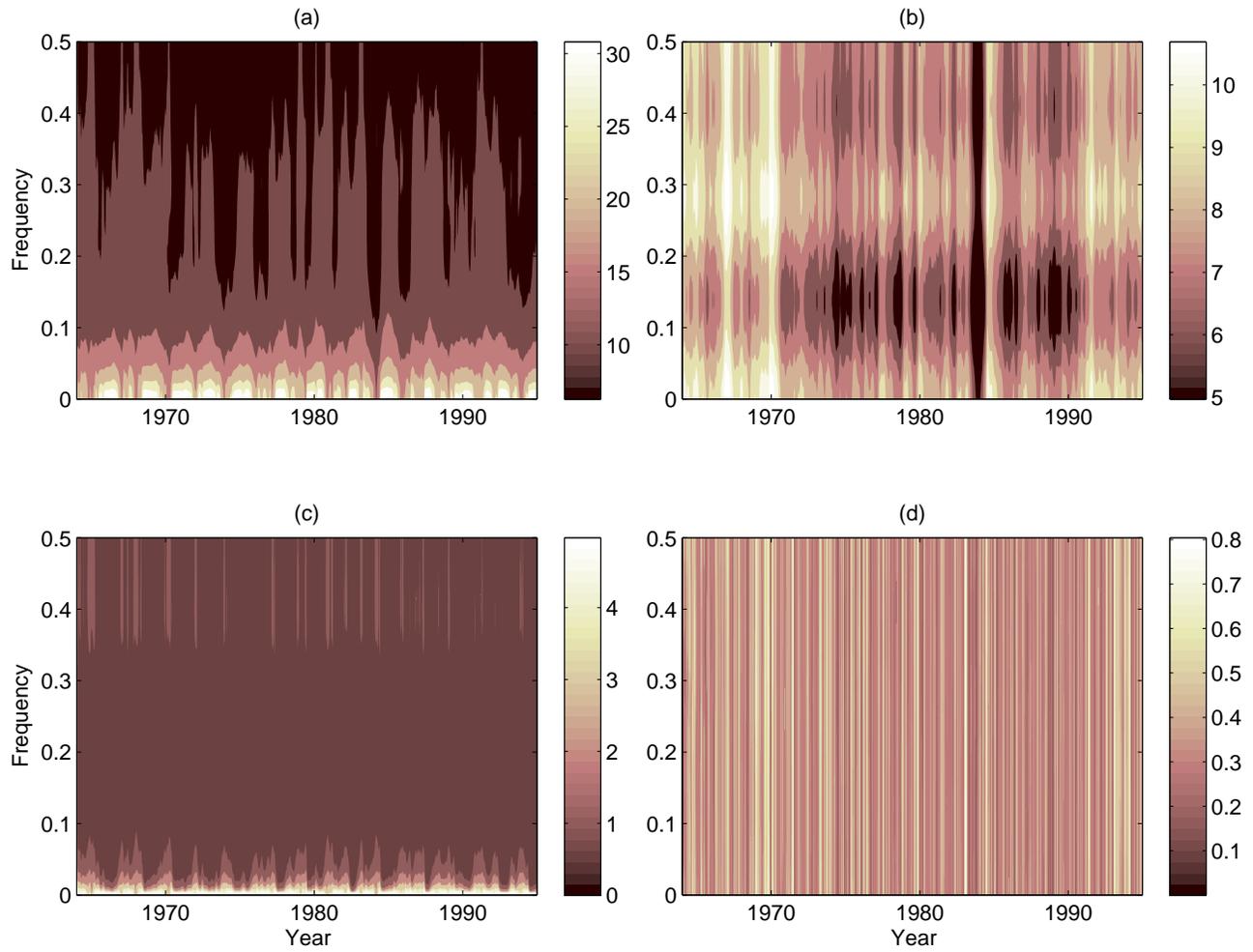}}
\caption{(a) and (c) display the posterior mean and standard deviation of time-frequency representations of the wind east/west component by fitting a TVAR(3) model. (b) and (d) display the posterior mean and standard deviation of time-frequency representations of the wind north/south component by fitting a TVAR(4) model.  \baselineskip=10pt } \label{fig:Windspec}
\end{figure}

\clearpage\pagebreak\newpage 

\clearpage\pagebreak\newpage 

\clearpage\pagebreak\newpage 

\section*{Appendix A: Sequential Updating and Smoothing  }
\begin{appendix}
\Appendix  
\renewcommand{\theequation}{A.\arabic{equation}}
\setcounter{equation}{0}
To complete the Bayesian estimation of the forward and backward PARCOR coefficients, as well as the time-varying innovation variances, we use dynamic linear models (DLMs) \citep[see,][]{west1997bayesian,prado2010time}. Specifically, we provide the details and formulas for analysis of $\{\alpha_{t,m}^{(m)}\}$ and $\{\sigma_{f,m,t}^{2}\}$. The analysis of $\{\beta_{t,m}^{(m)}\}$ and $\{\sigma_{b,m,t}^{2}\}$ follow similarly.       

For $t=1,\ldots, T$, given the values of $f_{t}^{(m-1)}$ and $b_{t-m}^{(m-1)}$ at stage $m$, recall (5) of Section~2.3 gives 
\begin{equation*}
f_{t}^{(m-1)}=\alpha_{t,m}^{(m)}b_{t-m}^{(m-1)} + f_{t}^{(m)},
\end{equation*}
with $f_{t}^{(m)}\sim N(0,\sigma_{f,m,t}^{2})$.  Modeling of $\{\alpha_{t,m}^{(m)}\}$ and $\{\sigma_{f,m,t}^{2}\}$ proceeds using (9) and (11):   
\begin{align*}
\alpha_{t,m}^{(m)} &= \alpha_{t-1,m}^{(m)} + \epsilon_{\alpha,m,t}, ~~~~ \epsilon_{\alpha,m,t}\sim N(0,w_{\alpha,m,t}),\\
\sigma_{f,m,t}^{2} &= \sigma_{f,m,t-1}^{2}(\delta_{f,m}/\eta_{f,m,t}),~~~~\eta_{f,m,t}\sim Beta(g_{f,m,t},h_{f,m,t}),
\end{align*}
with $w_{\alpha,m,t}=c_{f,m,t-1}(1-\gamma_{f,m})/\gamma_{f,m}$ \citep[see][Section 6.3]{west1997bayesian} and $c_{f,m,t-1}$ the scale parameter of the marginal $t$-distribution of $\alpha_{t-1,m}^{(m)}$ given the information up to time $t-1$.  Moreover, $g_{f,m,t}=\delta_{f,m}\cdot v_{f,m,t-1}/2$, and $h_{f,m,t}=(1-\delta_{f,m})\cdot v_{f,m,t-1}/2$ \citep[see][Section 10.8]{west1997bayesian}, where $v_{f,m,t-1}/2$ the shape parameter of the marginal gamma distribution of $\sigma_{f,m,t}^{-2}$ given the information up to time $t-1$ (see Section~\ref{appen:updating}). Then, given specified values for the two discount factors $\gamma_{f,m}$ and $\delta_{f,m}$, as well as conditional on assuming conjugate initial priors (\ref{eq:conjugate1}) and (\ref{eq:conjugate2}), we can specify the corresponding sequential updating and smoothing algorithms using DLM theory.    
  
\subsection{Sequential Updating}\label{appen:updating}
Using similar notation to \cite{west1997bayesian} and \cite{west1999evaluation}, we first sequentially update the joint posterior distributions of $p(\alpha_{t,m}^{(m)},\sigma_{f,m,t}^{-2}|D_{f,m,t})$ over $t=1,\ldots,T.$ Since the initial priors have the conjugate normal/gamma forms, $p(\alpha_{t,m}^{(m)},\sigma_{f,m,t}^{-2}|D_{f,m,t})$ also has the normal/gamma form. Therefore, the marginal posterior distribution of $\alpha_{t,m}^{(m)}$ is a $t$-distribution; i.e.,  $p(\alpha_{t,m}^{(m)}|D_{f,m,t})\sim T_{v_{f,m,t}}(\mu_{f,m,t},c_{f,m,t})$, with degrees of freedom $v_{f,m,t}$, location parameter $\mu_{f,m,t}$, and scale parameter $c_{f,m,t}$.  The marginal posterior distribution of $\sigma_{f,m,t}^{-2}$ is a gamma distribution $p(\sigma_{f,m,t}^{-2}|D_{f,m,t})\sim G(v_{f,m,t}/2,\kappa_{f,m,t}/2)$, with shape parameter $v_{f,m,t}/2$ and scale parameter $\kappa_{f,m,t}/2$. We summarize the sequential updating equations of parameters for $t=1,\ldots,T$, as follows:     
\begin{align*}
\mu_{f,m,t} &= \mu_{f,m,t-1} + z_{f,m,t}e_{f,m,t},\\
c_{f,m,t} &= (r_{f,m,t}-z_{f,m,t}^{2}q_{f,m,t})(s_{f,m,t}/s_{f,m,t-1}),
\end{align*}
and
\begin{align*}
v_{f,m,t} &= \delta_{f,m}v_{f,m,t-1}+1,\\
\kappa_{f,m,t} &= \delta_{f,m}\kappa_{f,m,t-1}+s_{f,m,t-1}e_{f,m,t}^{2}/q_{f,m,t},\\
s_{f,m,t} &= \kappa_{f,m,t}/v_{f,m,t}, 
\end{align*}
where
\begin{align*}
e_{f,m,t} &= f_{t}^{(m)}-\mu_{f,m,t-1}b_{t-m}^{(m-1)},\\
z_{f,m,t} &= r_{f,m,t}b_{t-m}^{(m-1)}/q_{f,m,t},\\
q_{f,m,t} &= r_{f,m,t}(b_{t-m}^{(m-1)})^2 + s_{f,m,t-1},\\
r_{f,m,t} &= c_{f,m,t-1} + w_{\alpha,m,t}, \\
w_{\alpha,m,t} &= c_{f,m,t-1}(1-\gamma_{f,m})/\gamma_{f,m}.
\end{align*}
Importantly, since $w_{\alpha,m,t} = c_{f,m,t-1}(1-\gamma_{f,m})/\gamma_{f,m}$, we can reduce $r_{f,m,t}$ to $r_{f,m,t}=c_{f,m,t-1}/\gamma_{f,m}$.    

\subsection{Smoothing}\label{appen:smoothing}
After the sequential updating process, we can use a retrospective approach to specify the smoothing joint distribution of   
$p(\alpha_{t,m}^{(m)},\sigma_{f,m,t}^{-2}|D_{f,m,T})$, for $t=1,\ldots,T$, given all the information up to time $T$ \citep{west1997bayesian,west1999evaluation}. We summarize the equations for the parameters of both marginal distributions, $p(\alpha_{t,m}^{(m)}|D_{f,m,T})\sim T_{v_{f,m,t|T}}(\mu_{f,m,t|T},c_{f,m,t|T})$, with degrees of freedom $v_{f,m,t|T}$, location parameter $\mu_{f,m,t|T}$, and scale parameter $c_{f,m,t|T}$.  Moreover, $p(\sigma_{f,m,t}^{-2}|D_{f,m,T})\sim G(v_{f,m,t|T}/2,\kappa_{f,m,t|T}/2)$, with shape parameter $v_{f,m,t|T}/2$ and scale parameter $\kappa_{f,m,t|T}/2$. It is important to note that when $t=T$, we have $\mu_{f,m,T|T}=\mu_{f,m,T}$, $c_{f,m,T|T}=c_{f,m,T}$, $v_{f,m,T|T}=v_{f,m,T}$, and $z_{f,m,T|T}=v_{z,m,T}$ from the results of sequential updating. Additionally, the point estimate of $\sigma_{f,m,t}^{2}$ at time $T$ is $s_{f,m,T|T}=s_{f,m,T}$. Then, for $t=T-1,\ldots,1$, we can summarize the equations as follows:     
\begin{align*}
\mu_{f,m,t|T} &= (1-\gamma_{f,m})\mu_{f,m,t} + \gamma_{f,m}\mu_{f,m,t+1|T},\\
c_{f,m,t|T} &= [(1-\gamma_{f,m})c_{f,m,t}+\gamma_{f,m}^{2}c_{f,m,t+1|T}](s_{f,m,t|T}/s_{f,m,t}),\\
v_{f,m,t|T} &= (1-\delta_{f,m})v_{f,m,t} +\delta_{f,m}v_{f,m,t+1|T},\\
1/s_{f,m,t|T} &= (1-\delta_{f,m})/s_{f,m,t} + \delta_{f,m}/s_{f,m,t|T},\\
\kappa_{f,m,t|T} &= v_{f,m,t|T}/s_{f,m,t|T}.
\end{align*}

\subsection{Algorithm for Fitting TVAR Models}\label{appen:algorithm}
We summarize the algorithm of our approach to fitting a TVAR($P$) model as follows:
\def\NoNumber#1{{\def\alglinenumber##1{}\State #1}\addtocounter{ALG@line}{-1}}

\begin{algorithmic}
\State {\bf Step 1.} Give a value for order $P$ and a set of values for $\{\gamma_{m},\delta_{m} \}$, for $m=1,\ldots,P$, as well as initial values of parameters at $t=0$.
\State {\bf Step 2.} For $t=1,\ldots,T$, set $f_{t}^{0}=b_{t}^{0}=x_{t}$.
\State {\bf Step 3.} Put $\{f_{t}^{0}\}$ and $\{b_{t}^{0}\}$ into (\ref{tvar_lattice1}) and then use sequential updating and smoothing to obtain $\{\widehat{\alpha}_{t,1}^{(1)}\}$, $\{\widehat{\sigma}_{f,1,t}\}$, and $\{f_{t}^{1}\}$.
\State {\bf Step 4.} Put $\{f_{t}^{0}\}$ and $\{b_{t}^{0}\}$ into (\ref{tvar_lattice2}) and then run  sequential updating and smoothing to obtain $\{\widehat{\beta}_{t,1}^{(1)}\}$, $\{\widehat{\sigma}_{b,1,t}\}$, and $\{b_{t}^{1}\}$.
\State {\bf Step 5.} Put $\{f_{t}^{1}\}$ and $\{b_{t}^{1}\}$ into (\ref{tvar_lattice1}) and (\ref{tvar_lattice2}) and then run  sequential updating and smoothing to obtain $\{\widehat{\alpha}_{t,2}^{(2)}\}$, $\{\widehat{\sigma}_{f,2,t}\}$, and $\{f_{t}^{2}\}$ as well as $\{\widehat{\beta}_{t,2}^{(2)}\}$, $\{\widehat{\sigma}_{b,2,t}\}$, and $\{b_{t}^{2}\}$. 
\State {\bf Step 6.} Repeat {\bf Step 5} until $\{\widehat{\alpha}_{t,P}^{(P)}\}$, $\{\widehat{\sigma}_{f,P,t}\}$, $\{\widehat{\beta}_{t,P}^{(P)}\}$,  and $\{\widehat{\sigma}_{b,P,t}\}$ are obtained.
\State {\bf Step 7.} Given the set of estimated values $\{\widehat{\alpha}_{t,m}^{(m)}\}$, $\{\widehat{\beta}_{t,m}^{(m)} \}$, for $m=1,\ldots,P$, use (\ref{tvar_levison1}) and (\ref{tvar_levison2}) iteratively to get the set of estimated $\{\widehat{a}_{t,m}^{(P)}\}$, $m=1,\ldots,P$, as well as set $\{\widehat{\sigma}_{t}^{2}=\widehat{\sigma}_{f,P,t}\}$. 

\end{algorithmic}

\end{appendix}

\pagebreak\newpage

\section*{Appendix B: Lattice Filter Structure  }
\begin{figure}
\begin{tabular}{c}
(a)\\
\includegraphics[width=1.1\textwidth]{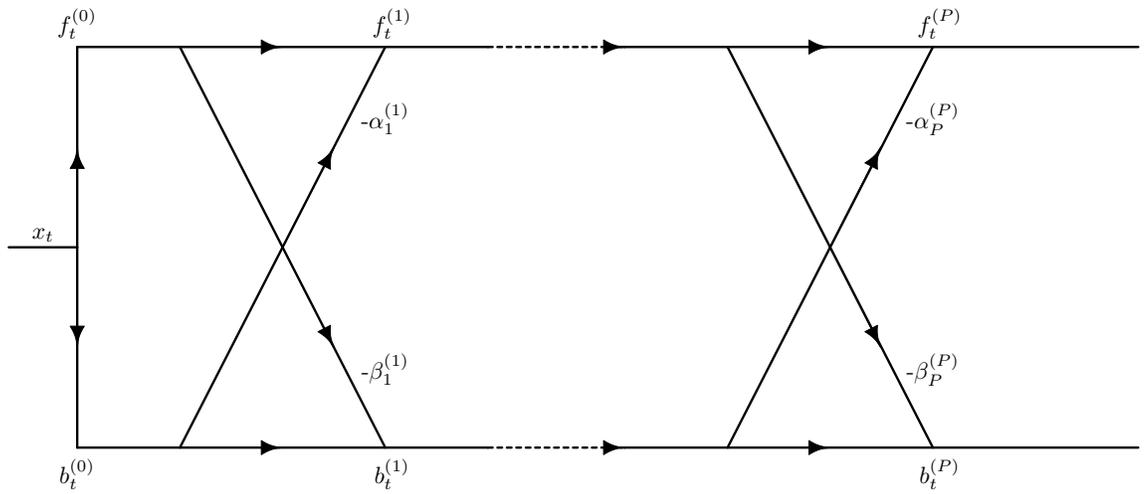}\\
(b)\\
\includegraphics[width=1.1\textwidth]{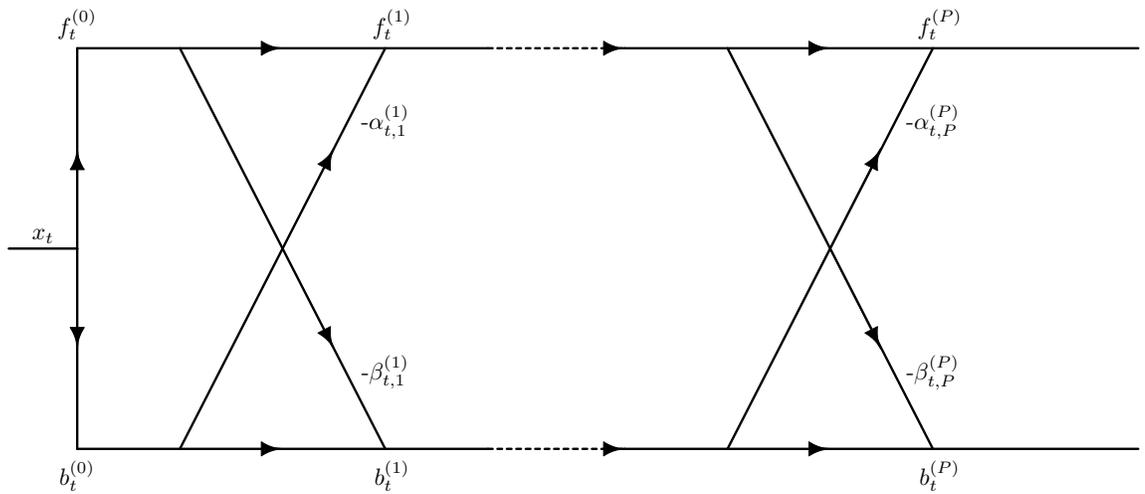}\\
\end{tabular}
\caption{(a) Graphical representation of the lattice filter for a stationary AR model. (b) Graphical representation of the lattice filter for a TVAR model. \baselineskip=10pt } \label{fig:LF}
\end{figure}

\begin{appendix}
\Appendix  
\renewcommand{\theequation}{B.\arabic{equation}}
\setcounter{equation}{0}
Figure~\ref{fig:LF}a illustrates the lattice structure for an AR($P$) model. Given $\alpha_{m}^{(m)}$ and $\beta_{m}^{(m)}$, $m=1,\ldots,P$, recursive use of (2) and (3) can produce forward and backward prediction errors for the forward and backward AR($m$) models (i.e., $f_{t}^{(m)}$ and $b_{t}^{(m)}$). Alternatively, Figure~\ref{fig:LF}b illustrates the lattice structure for a TVAR($P$) model. Different from Figure~\ref{fig:LF}a, the forward and backward PARCOR coefficients are time dependent. Given $\alpha_{t,m}^{(m)}$ and $\beta_{t,m}^{(m)}$, $m=1,\ldots,P$, this case presents recursive use of (5) and (6) to produce forward and backward prediction errors for the forward and backward TVAR($m$) model.  

\end{appendix}

\clearpage\pagebreak\newpage 

\clearpage\pagebreak\newpage 

\section*{Appendix C: Supplemental Figures for Case Studies}
\begin{appendix}
\Appendix  
\renewcommand{\theequation}{C.\arabic{equation}}
\setcounter{equation}{0}
This section contains supplemental figures associated with Section 4 of the main text (Case Studies).  These figures are not strictly necessary to illustrate the intended applications.  Nevertheless, we include them here for further potential scientific insight.

\begin{figure}
\centerline{
\includegraphics[width=0.8\textwidth,angle=-90]{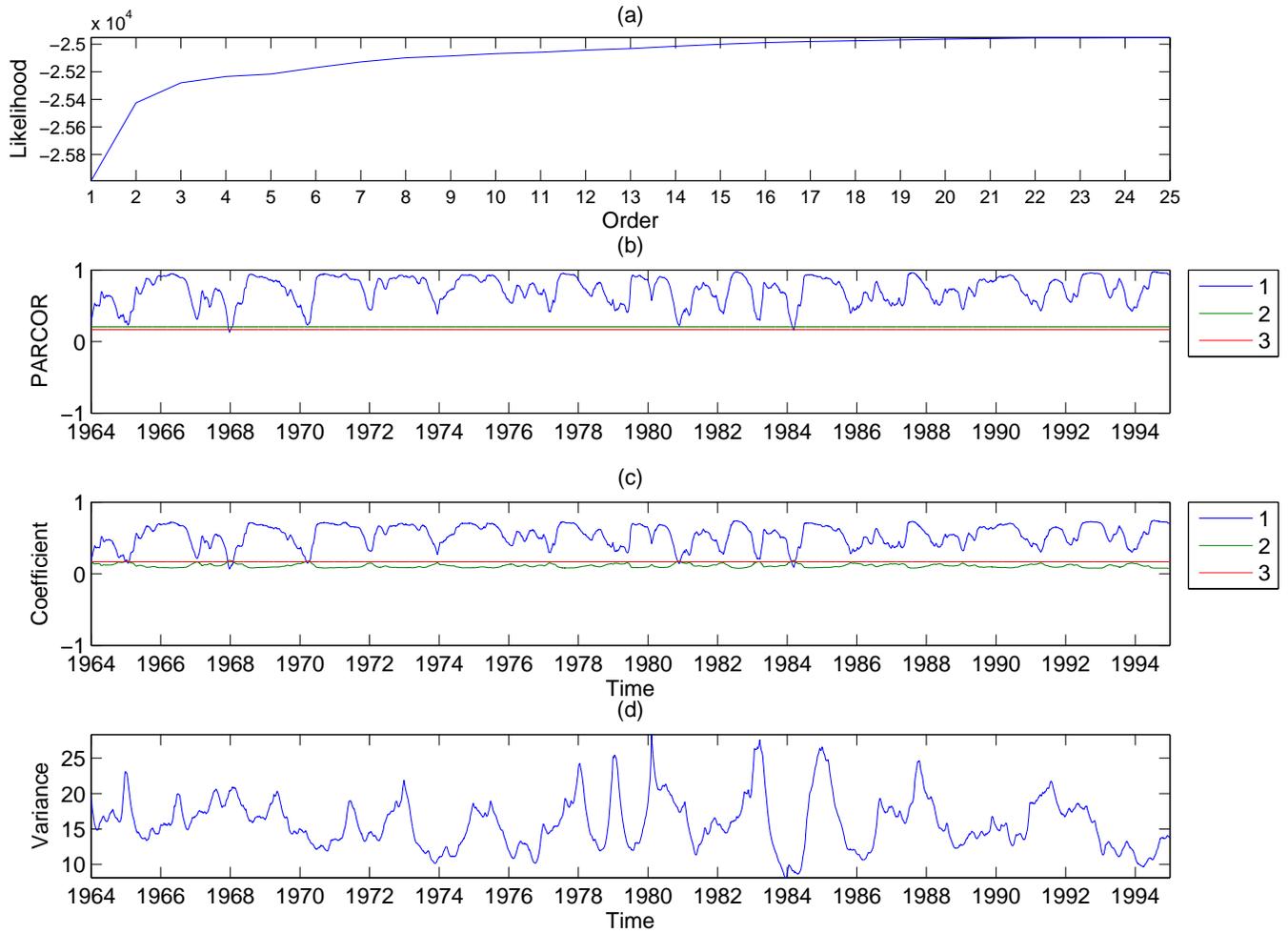}}
\caption{(a) shows the BLF-scree plot of the east/west component of the wind signal. (b) depicts the first three time-varying estimated PARCOR coefficients. (c) and (d) show the estimated time-varying coefficients and innovation variances of the TVAR(3) model.  \baselineskip=10pt } \label{fig:ewWind}
\end{figure}

\begin{figure}
\centerline{
\includegraphics[width=0.8\textwidth, angle = -90]{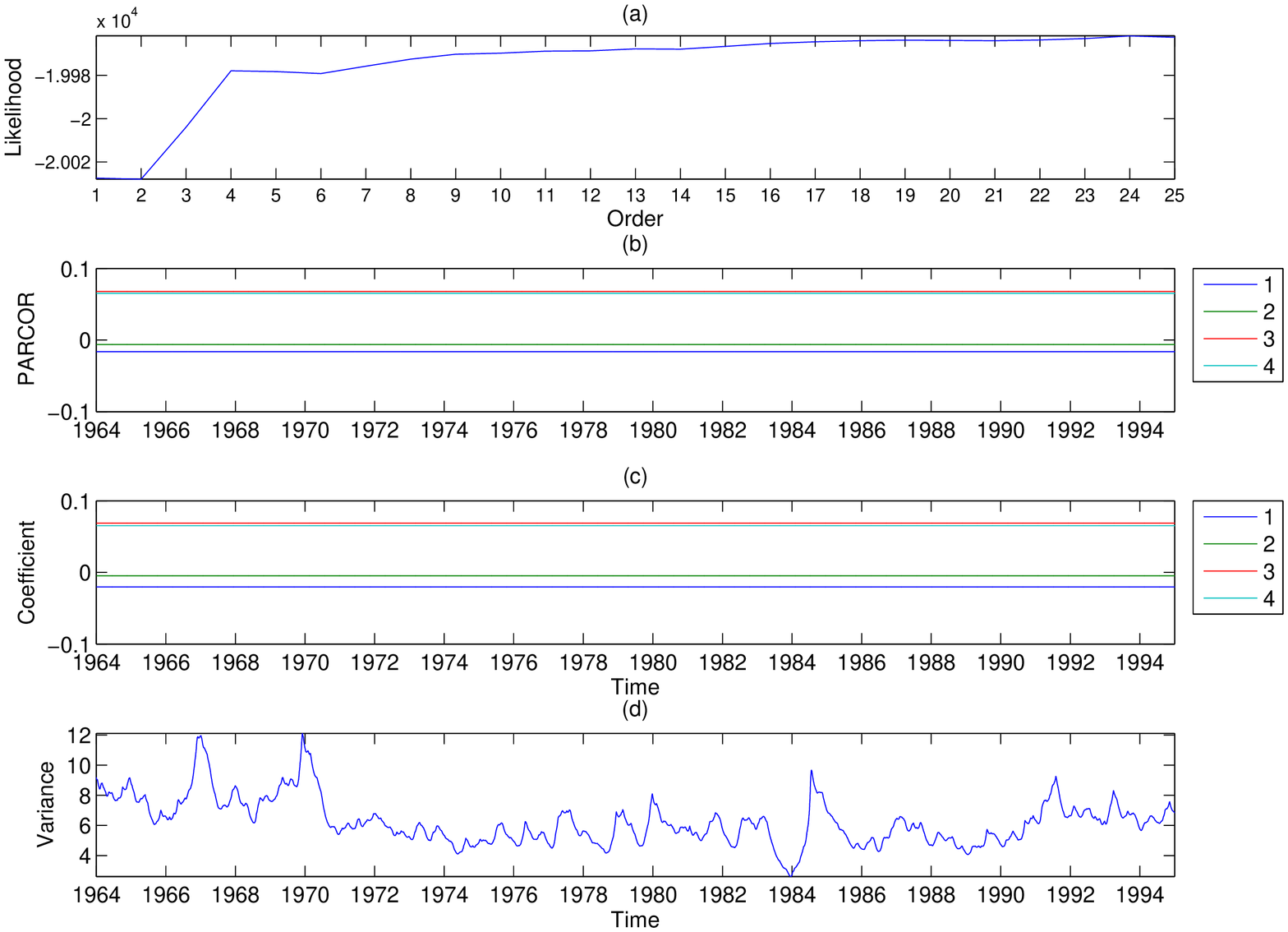}}
\caption{(a) shows the BLF-scree plot of the north/south component of the wind signal. (b) depicts the first four time-varying estimated PARCOR coefficients. (c) and (d) show the estimated time-varying coefficients and innovation variances of the TVAR(4).  \baselineskip=10pt } \label{fig:nsWind}
\end{figure}

\begin{figure}
\centerline{
\includegraphics[width=0.8\textwidth, angle = -90]{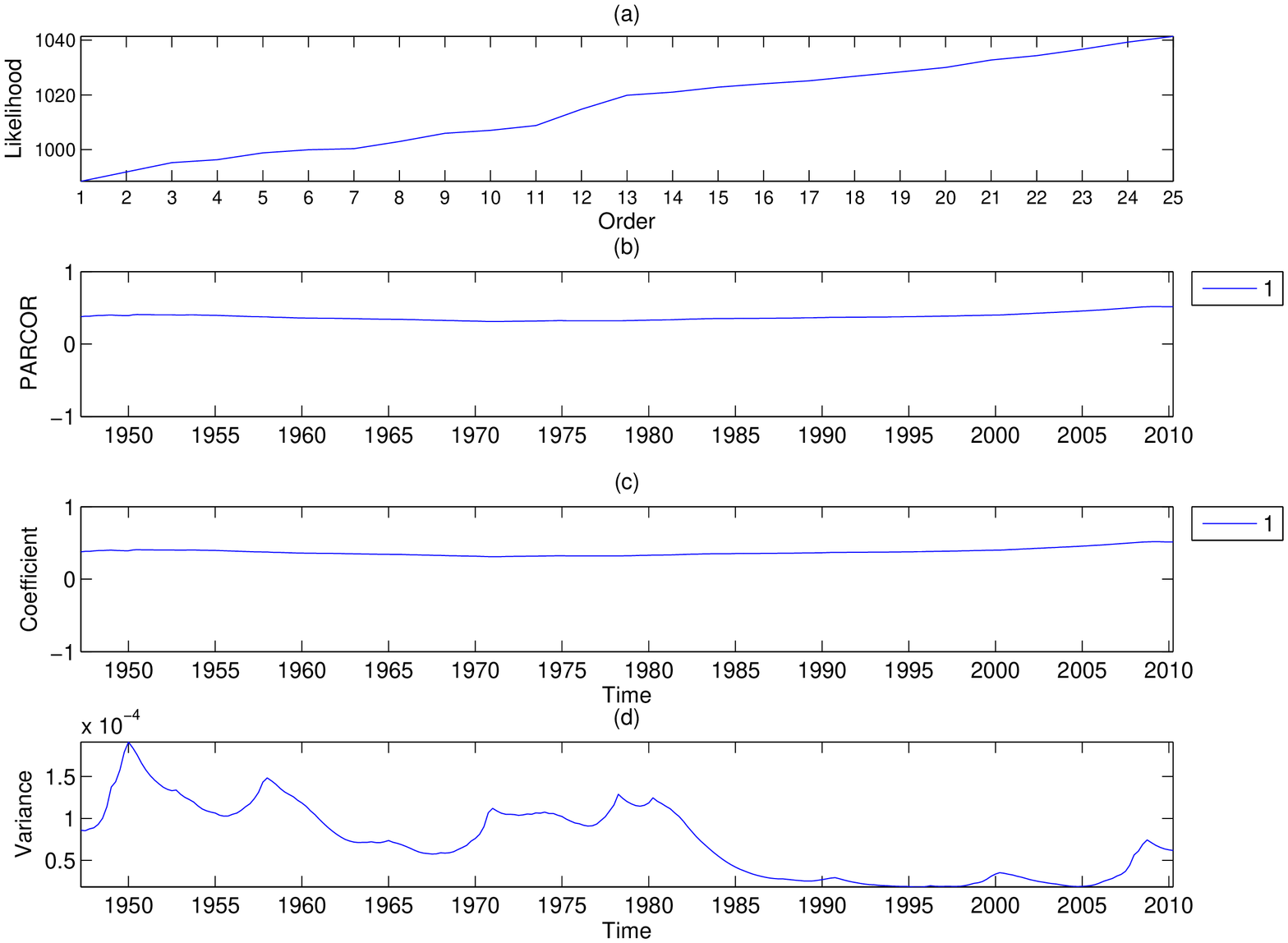}}
\caption{(a) shows the BLF-scree plot of the difference of logarithm GDP series. (b) depicts the first time-varying estimated PARCOR coefficients. (c) and (d) show the estimated time-varying coefficients and innovation variances of the time-varying AR(1).  \baselineskip=10pt } \label{fig:nsGDP}
\end{figure}

\begin{figure}
\centerline{
\includegraphics[width=0.8\textwidth, angle = -90]{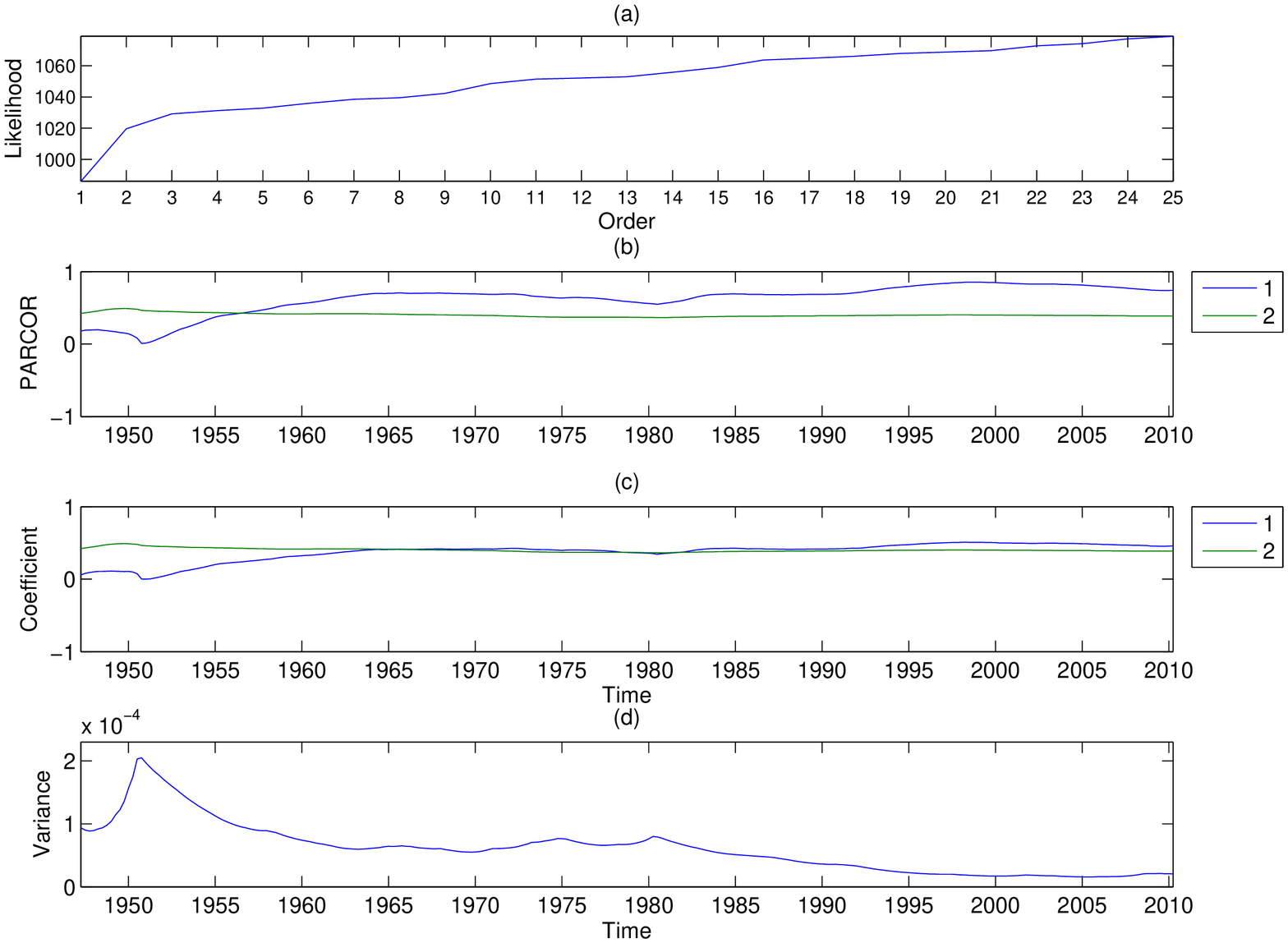}}
\caption{(a) shows the BLF-scree plot of the difference of logarithm consumption series. (b) depicts the first two time-varying estimated PARCOR coefficients. (c) and (d) show the estimated time-varying coefficients and innovation variances of the time-varying AR(2).  \baselineskip=10pt } \label{fig:nsConsump}
\end{figure}

Figures~\ref{fig:ewWind}a and \ref{fig:nsWind}a provide BLF-scree plots associated with the east/west and north/south components of the wind signal.  The first three time-varying estimated PARCOR coefficients of the  east/west and north/south components of the wind signal are provided in Figures~\ref{fig:ewWind}b and \ref{fig:nsWind}b.  The estimated time-varying coefficients of the TVAR(3) model for the east/west component are given in Figures~\ref{fig:ewWind}c and \ref{fig:ewWind}d whereas the estimated time-varying coefficients of the TVAR(4) model for the north/south component are given in Figures~\ref{fig:nsWind}c and \ref{fig:nsWind}d.  Finally, the BLF-scree plot, time-varying estimated PARCOR coefficients and time-varying innovation variances for the difference of the logarithm GDP and Consumptions series are provided in Figures~\ref{fig:nsGDP} and \ref{fig:nsConsump}. 

\end{appendix}

%
%
%
%
%


\clearpage\pagebreak\newpage 

\clearpage\pagebreak\newpage 

\clearpage\pagebreak\newpage 

\clearpage\pagebreak\newpage 

\clearpage\pagebreak\newpage 
\baselineskip=14pt \vskip 0mm\noindent
\bibliographystyle{jasa}
\bibliography{biblio}

\end{document}